\documentclass[12pt,preprint]{aastex}
\def\ltsima{$\; \buildrel < \over \sim \;$}
\def\simlt{\lower.5ex\hbox{\ltsima}}    
\def\gtsima{$\; \buildrel > \over \sim \;$}
\def\simgt{\lower.5ex\hbox{\gtsima}}    

\def\ref{\par\noindent\hangindent 20 pt}

\def\mincir{\ \raise -2.truept\hbox{\rlap{\hbox{$\sim$}}\raise5.truept 
\hbox{$<$}\ }}  %
\def\magcir{\ \raise -2.truept\hbox{\rlap{\hbox{$\sim$}}\raise5.truept %
\hbox{$>$}\ }}

\def\asec{$^{\prime\prime}$ }  

\begin{document}

\title{The cosmic evolution of quasar host galaxies}

\author{Renato Falomo}
\affil{INAF -- Osservatorio Astronomico di Padova, Vicolo dell'Osservatorio 5, 
35122 Padova, Italy}
\email{falomo@pd.astro.it}

\author{Jari K. Kotilainen}
\affil{Tuorla Observatory, University of Turku, V\"ais\"al\"antie 20, 
FIN--21500 Piikki\"o, Finland}
\email{jarkot@astro.utu.fi}

\author{Claudio Pagani }
\affil{Universit\`a dell'Insubria, via Valleggio 11, 22100 Como, Italy}
\email{pagani@mib.infn.it}

\author{Riccardo Scarpa }
\affil{European Southern Observatory, 3107 Alonso de Cordova, 
Santiago, Chile. }
\email{ rscarpa@eso.org}

\author{Aldo Treves}
\affil{Universit\`a dell'Insubria, via Valleggio 11, 22100 Como, Italy}
\email{treves@mib.infn.it}

\begin{abstract}

We present the results of a near-infrared imaging study of the 
host galaxies of 17 quasars in the redshift range 1$<z<$2.   
The observations were carried 
out at the ESO VLT UT1 8m telescope under  excellent seeing 
conditions ($\sim$ 0\farcs4). The sample includes radio-loud (RLQ) 
and radio-quiet (RQQ) quasars with similar distribution of redshift  
and optical luminosity.  For all the observed objects but one we have 
been able to derive the global properties of the surrounding nebulosity. 
The host galaxies of both types of quasars appear to follow the expected 
trend in luminosity of massive ellipticals undergoing simple 
passive evolution.  However, we find a systematic difference by a 
factor $\sim$2 in the host luminosity between RLQs and RQQs 
($<M_K>_{RLQ}$(host) =--27.55 $\pm$ 0.12 and 
$<M_K>_{RQQ}$(host) = --26.83 $\pm$ 0.25). Comparison with other samples 
of quasar hosts at similar and lower redshift indicates  that the 
difference in the host luminosity between RLQs and RQQs remains the same 
from z = 2 to the present epoch. 
No significant correlation is found between the nuclear and the 
host luminosities.  Assuming that the host luminosity is proportional to 
the black hole mass, as observed in nearby massive spheroids, 
these quasars emit at very different levels (spread $\sim$ 1.5dex) with respect 
to their Eddington luminosity and with the same distribution for 
RLQs and RQQs. Apart from a factor of $\sim$2 difference in luminosity, 
the hosts of RLQs and RQQs of comparable nuclear luminosity appear to 
follow the same cosmic evolution as massive inactive spheroids.  
All together, our results support a view where nuclear activity can occur 
in all luminous ellipticals without producing a significant change in 
their global properties and evolution.  Quasar hosts appear to be already 
well formed at z $\sim$2, in disagreement with the predictions of 
models for the joint formation and evolution of galaxies and active nuclei 
based on the hierarchical structure formation scenario.

\end{abstract}


\section{Introduction}

In the local Universe (z \ltsima 0.3) images of powerful 
active galactic nuclei (AGN), i.e. quasars, clearly show that they are 
hosted by massive galaxies.  Ground-based imaging (e.g. 
McLeod \& Rieke 1994; 
\citealp{taylor96}; \citealp{kotifal00}; \citealp{perci00}) have 
been complemented by higher resolution data obtained by the 
Hubble Space Telescope (HST; e.g. \citealp{disney95};  \citealp{B97}; 
Hooper, Impey \& Foltz 1997; \citealp{boyce98}; \citealp{hutch99}; 
\citealp{hamilton02}; Dunlop et al. 2003; Pagani, Falomo, \& Treves 2003) 
and clearly indicate that 
the majority of quasar hosts are massive galaxies dominated by the spheroidal component. 
This result is consistent with the recent discovery that nearby 
massive spheroids (ellipticals and bulges of early type spirals) have 
an inactive supermassive black hole (BH) in their centers 
(see e.g. \citealp{ferrarese02} for a recent review).  These observations depict 
an evolutionary scenario where nuclear activity may be a common phenomenon 
during the lifetime of a galaxy with recurrent accretion episodes, 
and that the emitted nuclear power depends on the mass of the system. 
Powerful nuclear (quasar-like) activity is in fact 
only found in the most luminous (massive) galaxies (Hamilton et al. 2002;
\citealp{falomo03}; Kauffmann et al. 2003.

While radio-loud quasars (RLQ) are exclusively hosted by ellipticals 
exceeding the characteristic galaxy luminosity L* \citealp{mob93} by $\sim$2-3 mag 
and similar to the brightest cluster galaxies, 
radio-quiet quasars (RQQ) are found both in ellipticals and in 
early type spirals (Taylor et al. 1996; Bahcall et al. 1997). 
However, there is  evidence \cite{dunlop03} that at high 
nuclear luminosities also RQQs are hosted mainly in elliptical galaxies.  
There is also some indication at low redshift that the hosts of RLQs 
are systematically more luminous than those of RQQs 
(\citealp{veron90}; \citealp{B97}, \citealp{dunlop03}).

The strong cosmological evolution of the quasar population 
(\citealp{dunlop90}; Warren, Hewett, \& Osmer 1994;
\citealp{boyle01}) is similar to 
the evolution of the star formation history in the Universe 
(Madau, Pozzetti, \& Dickinson 1998; 
\citealp{franc99}; \citealp{steidel99}) and to the 
number density of radio galaxies (RG; \citealp{boyle98}. This may represent 
the overall effect of a fundamental link between the formation of 
massive galaxies and the formation and fuel ling of their nuclei, 
consistently with the finding of supermassive BHs in the nuclei of 
nearby inactive galaxies (e.g. \citealp{kor01} and references therein).

Deep high spatial resolution HST images of distant galaxies 
(e.g. \citealp{abraham96}; Koo et al. 1996; \citealp{lefevre}) have begun 
to provide data able  to trace the galaxy formation, while very little 
is still known about the evolution of distant quasar hosts. 
In the present epoch, quasar activity is a rare event in galaxies while it 
was a more common phenomenon at an earlier epoch (z $\sim$2--3) when the 
age of the Universe was only a few Gyr. This dramatic  evolution of 
quasars must thus be connected with the formation and evolution of 
massive spheroids \cite{franc99}. Understanding how  the properties of 
the galaxies hosting quasars change with the cosmic time is therefore 
a fundamental step to investigate the link between evolution of the 
galaxies and nuclear activity. In particular, it is of great importance 
to probe the host properties close to (and possibly beyond) the peak of 
quasar activity.

The detection of the host galaxies and the characterization of 
their properties are more and more difficult as one moves to 
higher redshift.  This is because the surrounding nebulosity becomes 
rapidly very faint compared to the nuclear source. This problem  is 
critical when studying high luminosity AGN. In order to cope with 
these severe limitations, it is imperative to obtain images of the 
targets with the highest possible spatial resolution and sensitivity. 
Moreover, a well defined point spread function (PSF) is crucial when 
modeling the image of the object. These requirements are seldom matched by using 
ground-based medium size (4m class) telescopes even under good 
seeing conditions. They are partially satisfied by HST that 
certainly has  a superbly narrow PSF but, due to its 
small aperture, has a relatively small throughput.

In spite of these severe difficulties, a number of studies have already 
been presented for quasar hosts at z $>$ 1 and in some cases extended 
emission has been reported for quasars even at z \gtsima 2 
(e.g. \citealp{heckman}; \citealp{lehnert92};  \citealp{lowenthal}; 
Aretxaga, Terlevich \&  Boyle 1998; \citealp{hutch98}, 1999; \citealp{lehnert99}; San).
However, the results of most of these studies are limited by modest seeing and/or 
image deepness. A further complication may arise from the 
contamination, inside the broad band observed, of line emission that could originate 
in spatially extended regions of gas around the nucleus of the quasar.
Moreover, the usually small number of objects investigated 
and the non-homogeneous data sets have failed to provide an unambiguous 
view of the evolution of quasar hosts and of the differences between 
RLQ and RQQ hosts.

The most systematic study until now on high redshift quasar host galaxies, 
based on HST NICMOS observations, has recently been presented 
by \citealp{Kuk01}.  They derived the host galaxy luminosities for a 
small sample of both RLQs and RQQs at z $\sim$1 and z$\sim$2 and compared 
them with the properties of quasar hosts at lower redshift. They found 
that the evolution of RLQ hosts is roughly consistent with that of 
massive ellipticals undergoing passive evolution while the luminosity of 
RQQ hosts remains nearly constant.  In neither case is there a 
significant drop in the host mass as would be expected in the models 
of hierarchical formation of massive ellipticals \cite{kauffmann00}. 
\citealp{Kuk01} also find evidence for a systematic gap between RLQ and 
RQQ host luminosity which appears to increase with the redshift.
 
Taking advantage of both the excellent PSF and the high throughput of the 
8m Very Large Telescope (VLT), we have carried out a program to image and 
to characterize the host galaxies of quasars in the redshift range 
1 $<$ z $<$ 2. The first results of this program for three RLQs at 
z $\sim$1.5 were reported in Falomo, Kotilainen \& Treves 2001, 
hereafter \citealp{FKT01}. In this paper 
we present the complete results of this program for all the 17 observed 
RLQs and RQQs.  In section 2 we describe our observed sample while in 
section 3 we report the observations and describe the data analysis. 
In section 4 we give our results for the observed quasars and compare 
them with the host luminosities of quasars derived from other samples. 
Finally, the cosmic evolution of RLQ and RQQ host galaxies and 
the relationship between host and nuclear luminosities are discussed 
in section 5.  For consistency with previous studies, we adopt 
Hubble constant H$_0$ = 50 km s$^{-1}$ Mpc$^{-1}$ and $\Omega$ = 0  
throughout this paper.

\section{The sample}

The observed targets were extracted from the list of objects reported 
in the catalogue of \  \citealp{veron01} \ requiring: 1.0 $<$ z $<$ 2.0,  
-25.5 $< M_B <$ -28 and --60$^\circ$ $<\delta <$ -8$^\circ$, 
and having sufficiently bright stars within the observed field of view 
($\sim$2 arcmin) in order to allow a reliable characterization of the PSF. 
We included both RLQs and RQQs in order to investigate the difference 
between the host galaxies of the two types of quasar. We considered a 
sample of 26 quasars that are evenly distributed in redshift and 
optical luminosity. An equal number of RLQs and RQQs were taken, 
matching their redshift and optical luminosity distributions. In total, 
14 of these sources were imaged during the two campaigns reported here 
(see Table 1), in addition to the three objects described in \citealp{FKT01}.  
Fig. 1 shows  the distribution of the observed quasars in 
the redshift--optical luminosity plane compared with all the quasars in 
the \citealp{veron01} catalogue. The average redshift of the observed quasars 
is $<z>$ = 1.51 $\pm$ 0.16  for  ten RLQs and $<z>$ = 1.52 $\pm$ 0.16 
for  seven RQQs. The average luminosity of the observed quasars is 
$< M_B >$= -26.75 $\pm$ 0.73 (rms) and $< M_B>$= -26.70 $\pm$ 0.84 (rms) 
for the RLQs and RQQs, respectively. Our observed samples are thus 
well matched, and lie toward the high luminosity end of the quasar in the 
\citealp{veron01} catalogue.

To perform the comparison between the hosts of RLQs and RQQs, it is important
to ensure that the RLQs are genuinely radio-loud 
(P(5 GHz) $>$ 10$^{25}$ W/Hz/sr), and that the RQQs are genuinely radio-quiet 
(P(5 GHz) $<$ 10$^{24.5}$ W/Hz/sr). 
For our sample of RLQ the 
average 5 GHz (6 cm) radio luminosity is 
$<log P(5 GHz)>$ (W/Hz/sr) = 27.18 $\pm$ 0.40 (rms). Note that even the 
radio-faintest of the RLQs ( Log P(5 GHz) = 26.3 W/Hz/sr) is well beyond the 
threshold for radio loud objects.
No radio data are available for the RQQs in the sample both from the QSO catalogue and from NED.

Of the 10 RLQs, eight are steep spectrum radio quasars  (SSRQ; $\alpha$ $<$ 0) and two 
flat spectrum radio quasars  (FSRQ; $\alpha$ $>$ 0). The average 6-11 cm radio spectral index of the 
observed RLQs is $<\alpha(6-11 cm)>$ = -0.18 $\pm$ 0.64 (rms). 
Given the small fraction of FSRQ we assume that beaming effects for the nuclear emission are 
negligible

\section{Observations}

Deep images of the quasars in the $H$- or $K$-band were obtained using 
the near-infrared (NIR) ISAAC camera \cite{cuby}, mounted on the first 
8m unit telescope (UT1, Antu) of VLT at 
the European Southern Observatory (ESO) in Paranal, (Chile).  
At the redshift of the objects, the observed bands correspond to rest frame 
$R$- and $I$-bands where most of the studies for low redshift objects 
have been performed.  The Short Wavelength (SW) arm of ISAAC is equipped 
with a 1024 x 1024 px Hawaii Rockwell array, with a pixel scale of 
0\farcs147 px$^{-1}$, giving a field of view of $\sim$150 x 150 arcsec. 
The observations were performed in service mode in the period 2001 June 
to 2002 May.

A detailed journal of the observations is given in Table 1.  The seeing, 
as derived from the median full with half maximum (FWHM) size of the image 
of stars in each frame,  
was consistently excellent during all observations, ranging from  
$\sim$0\farcs32 to $\sim$0\farcs58 (average $<$FWHM $>$= 0\farcs41; median 0\farcs39).

Total integration times were $\sim$60 minutes and $\sim$30 minutes for 
targets above and below z = 1.4, respectively. In order to maintain 
the stability of the observing conditions (in particular of the seeing) 
during the  integration time, we typically obtained pairs of images of 
$\sim$30 minutes each to reach the 60 minutes of total integration. 
The images were secured using a jitter procedure and individual exposures of 
2 minutes per frame. The jittered observations were controlled by an 
automatic template (see \citealp{cuby}), which produced a set of 
frames slightly offset in telescope position from the starting point. 
The observed positions were randomly generated within a box of 
10 x 10 arcsec centered on the first pointing. Each frame was flat-fielded 
and sky-subtracted and the final image was produced for each quasar 
by co-adding these frames. 
Data reduction was performed by the ESO pipeline for jitter imaging 
data \cite{devillard99}. The normalized flat field was obtained by 
subtracting ON and OFF images of the illuminated dome, 
after interpolating over bad pixels. Sky subtraction was done by 
median averaging sky frames from the 10 frames nearest in time.  
The reduced frames were aligned to sub-pixel accuracy using 
a fast object detection algorithm, and co-added after removing  
spurious pixel values. Photometric calibration was performed using 
standard stars observed during the same night. The estimated 
internal photometric accuracy is $\pm$0.03 mag.  We have also performed 
an additional check of the photometric calibration based on field stars 
that have NIR magnitudes in the 2MASS point source catalog. 
 We find 3-4 stars in the  fields of PKS 2210--25, PKS 2227--08 and 
PKS 1511--10. The agreement between 2MASS and our photometry is in 
all cases within 0.1 magnitudes.
For three objects in the sample we found 
previous NIR photometry published in the literature (Francis, Whiting \& Webster 2000):
the photometry obtained through a 5 arcsec aperture (
 K = 14.70 for PKS 1511-100,  K = 15.65 for 
PKS 2210-257, and  K = 15.06 for PKS 2227-088) differs by 0.1--0.7 magnitudes 
with ours (cf. Table 2) indicating a moderate NIR nuclear variability.

The use of  the $H$- and $K$-bands combined with  observing in the 
1.3 $< z <$ 1.8 redshift interval implies that we are sampling a rest frame 
interval of $\sim$300 \AA \  between 6500 and 8900 \AA, depending on 
the redshift of the object. In this region the only relevant strong 
emission line is H$\alpha$ at 6563 \AA. The averaged  rest frame wavelength sampled for 
the seven RLQs and seven RQQs presented here varies between 
7000 and 8900 \AA ~and exclude this emission line. Note, however, 
that for the three z $\sim$1.5 RLQs studied by \citealp{FKT01} in 
the $H$-band, some contamination from the H$\alpha$ line may be present.

\section{Data analysis}

To detect and characterize the properties of the host galaxies of quasars, the key factors 
are the apparent nucleus-to-host magnitude ratio and the seeing (shape of the PSF).
While the total magnitudes of the hosts are relatively easily determined,
the scale-lengths are less well constrained.
The most critical part of the analysis is to perform a detailed study of the PSF 
for each frame.  In particular, it is important to have a sufficient number 
of reference stars distributed over the field of view in order to account 
for any possible positional dependence of the PSF. Moreover, it is 
essential to have at least one sufficiently bright star in the field to 
allow a reliable evaluation of the shape of the faint wing of the PSF, 
against which most of the signal from the surrounding nebulosity will 
be detected. 

The relatively large field of view of ISAAC ($\sim$2\farcm5) and 
the constraint on the quasar selection to have at least one bright star in 
the field of view, allowed us to reach this goal and thus to perform 
a trustworthy characterization of the PSF. For each field, we analyzed 
the shape of all stellar profiles and constructed a composite PSF, 
the brightness profile of which extends down to 
$\mu_K$ $\sim$24.5 mag arcsec$^{-2}$. This guarantees a reliable 
comparison between the luminosity profiles of the quasars and of the 
stars without requiring blind extrapolation of the PSF at large radii 
(faint fluxes) that could produce spurious results.  The shape of the 
PSF profile was found to be symmetric (ellipticity less than few percent) 
and very stable across the field of the images. 
The differences of FWHM of the stars in each frame were typically 
less than a few percent, while no significant difference was found among 
their radial brightness profiles. In Fig. 2 we show an example of 
the azimuthally averaged radial profiles of stars used to construct a PSF, 
together with the overall deviations from the used PSF model of 
individual stellar profiles. 

In all our objects the emission from nuclear source is clearly dominant with respect 
to the light from the extended surrounding nebulosity.
A first  indication of the presence of a surrounding nebulosity can be obtained 
after the subtraction of a scaled PSF. 
However, visual inspection of these PSF subtracted images  allow one to 
see residual emission only for the objects where the contrast between 
nucleus and host galaxy is relatively low
(see the examples reported in Figure 3). 
In other cases the high contrast  of the components (nucleus and host) of the 
objects prevents to clearly visualize the extended nebulosity above the 
signal-to-noise per pixel of the images. 

In order to  improve 
the S/N of the data and the capability to detect the faint signal from the host galaxies 
we have therefore computed for each quasar  the azimuthally averaged fluxes as a 
function of the distance from the nucleus, excluding any region around 
the quasars contaminated by companion objects. These companions are easily recognized from 
the original and PSF subtracted images since they are in all cases rather compact 
features covering an area of few tenths of arcsec in the image. 
To perform this cleaning we substituted the  area contaminated by possible 
companions with the corresponding one  
in the image which is symmetric with respect to the center of the target. 
In this  way we avoid to remove also the emission from the underlying galaxy and assume that it is 
essentially regular.
With this procedure we obtained  
the radial luminosity profile out to a radius where the signal 
becomes indistinguishable from the background noise.  For our observations, 
this level corresponds to $\mu$(K) $\sim$23--24 mag arcsec$^{-2}$, 
typically reached at $\sim$2$''$--3$''$ distance from the nucleus. 
This procedure allowed us to significantly improve the S/N at the faint 
fluxes where the signal from the host galaxy become detectable  with respect to 
that from the unresolved nuclear source. 

A straightforward comparison of the average radial brightness  profile with that of  
the proper  PSF gives us 
a first indication of the amount of the extended emission. 
Detailed modeling of the luminosity profile was then carried out using 
an iterative least-squares fit to the observed profile, assuming a 
combination of a point source (modelled by the PSF) and  an elliptical galaxy 
described by a de Vaucouleurs r$^{1/4}$ law, convolved with the proper PSF. 

We also attempted a fit using a pure exponential disk model for the 
host galaxy. Note, however, that the small extent of the hosts and 
the dominance of the nuclear emission for the observed targets make it 
very difficult to discriminate between the two models. Nevertheless, 
in all cases where the object is well resolved, we find that the 
elliptical model  yields a better fit than a disk model.  
Therefore, based on this, and consistently with the properties of 
lower redshift RLQs, we have assumed the elliptical model for 
the determination of the host galaxy properties in the following discussion. 
If a disk model were assumed, the luminosities of the hosts 
would systematically become $\sim$0.3 mag fainter.  This difference 
does not affect the  main conclusions of this study.

With the applied procedure we can derive the luminosity and the 
scale-length of the host galaxies and the luminosity of the nuclei. 
We have estimated the accuracy of the decomposition to derive the 
host parameters, taking into account the uncertainty of the observed profile 
(which is limited mainly by the signal-to-noise at the faintest flux levels) 
and the accuracy of the PSF shape.  We assumed the uncertainty in the 
derived parameters  for a variation of $\chi^2_\nu$ = 2.7 (for 2 degrees of freedom). 
While the total magnitude of the host galaxy can be derived with a typical 
internal error of 0.2 - 0.3 magnitudes, the scale-length is often poorly 
constrained.  This depends on the degeneracy that occurs between two model parameters: 
the effective radius $r_e$ and the surface brightness $\mu_e$. In fact for a 
given value of the total magnitude of the host, various pairs of 
$r_e$ and $\mu_e$ can fit the data without a significant difference in 
the $\chi^2_\nu$ value (see also \citealp{abraham92}; \citealp{taylor96}; 
\citealp{dunlop03} and \citealp{pagani03} 
for further discussion on this issue).

\section{Results}

In Fig. 4 we report for each quasar the observed radial brightness profile 
and the best fit  using the elliptical galaxy model and the 
procedure described above. The parameters of the best fit, together with 
their estimated uncertainty, are given in Table 2. For all quasars except one 
(HE 0935-1001) we find significant systematic deviations of the 
radial profile with respect to its proper PSF. This is quantified in Table 2 
by the ratio of the reduced $\chi^2_\nu$ value of the best fit with 
that obtained from the fit excluding the galaxy component, 
i.e. considering only the PSF. 

In Table 3 we give the absolute magnitudes and the effective radii for 
each quasar host, including the three RLQs analyzed in \citealp{FKT01}. 
The absolute magnitudes of the host galaxies have been K-corrected  using 
the optical-NIR evolutionary synthesis model for elliptical galaxies 
\cite{poggianti}. For the nuclear magnitudes we applied a correction 
$\Delta m$ = -2.5($\alpha$+1)Log(1+z). No correction for 
Galactic extinction was applied since it is negligible in the observed 
NIR bands. Moreover, to make the results homogeneous we transformed 
the $H$-band magnitudes to the $K$-band, assuming an intrinsic color 
$H$-$K$ = 0.2 typical of ellipticals and $H$-$K$ = 1 for the nucleus. 

\subsection{Host galaxies of RLQ and RQQ between z = 1 and z = 2}

In the following, we describe the properties of our full sample of 
16 resolved quasars. The average absolute $K$-band magnitude of the 
host galaxies is $<M_K>$(host) = {--27.55 $\pm$ 0.12}  and 
$<M_K>$(host) = {--26.83 $\pm$ 0.25} for the RLQs and RQQs, respectively.
The average absolute $K$-band magnitude of the nuclei after taking into 
account the above mentioned K- and color corrections 
is $<M_K>$(nucleus) = --30.94 $\pm$ 1.2 and 
$<M_K>$(nucleus) = --30.20 $\pm$ 1.2 for the RLQs and RQQs, 
respectively.  

We plot in Figure 5 the absolute $K$-band magnitude of the quasar host galaxies 
versus the redshift.  All the observed quasars have host galaxies with 
luminosity ranging between M$^*$ and M$^*$-2, where 
M$^*$(K) = --25.2 (Mobasher, Sharples, \& Ellis 1993) is the characteristic luminosity of the 
Schechter luminosity function for elliptical galaxies. 
For comparison, we also report in Figure 5 the absolute magnitudes of four RLQs 
and five RQQs at z $\sim$1.9 \cite{Kuk01} and three RQQs at 
z $\sim$1.8 \cite{ridgway01}, derived from HST NICMOS imaging studies. 
Note that the objects in these samples cover a large range in nuclear luminosity, 
and are on average less luminous than those in the sample considered here.   
In order to treat these literature data homogeneously, we have 
considered the published apparent magnitudes in the $J$ and $H$-band 
(HST filters F110M and F165M) and 
transformed them to M$_K$ following our procedure (K-correction, cosmology and 
color correction). In particular, we converted the $H$-band magnitudes in 
Table 2 of \citealp{Kuk01}, that are in the HST magnitude system 
(Kukula, private communication) into the standard IR Johnson system.  To do this, 
we computed synthetic color transformations from the F110M and F160M 
HST filters to $J$- and $H$-band assuming the input spectrum of an 
elliptical galaxy (\citealp{kinney96}) and the passband curves of the filters. 
This yields a correction of $\sim$0.5 and $\sim$0.2 mag for the $J$- and 
$H$-bands, respectively. For the three RQQs observed by \citealp{ridgway01}, 
we converted their published $H$-band fluxes into $H$-band magnitudes and 
then applied the corrections to the aperture magnitudes (their Tables 3 and 5) 
to obtain the total magnitudes of the hosts. Note that all the HST data show 
a substantially larger scatter than our VLT data. In particular, four out of 
the five RQQs observed by \citealp{Kuk01} lie above M$^*$ while all three 
RQQs observed by \citealp{ridgway01} are at or below M$^*$. The reason for 
this larger scatter is unclear but could be partially related to non-homogeneous 
data analysis. 
While the host parameters in this work and in \citealp{Kuk01} are derived 
using (1D or 2D) modelling of the brightness distribution of the sources, 
the measurements of \citealp{ridgway01} 
are obtained from aperture fluxes and, in spite of the applied corrections, 
could still underestimate the host galaxy luminosity. On the other hand 
the nuclei of two of the three objects studied by \citealp{ridgway01} 
are about 4 magnitudes fainter than the average luminosity of the 
objects in our sample. This may suggest  some dependence of the 
host galaxy luminosity on the nuclear luminosity. The available data 
are, however, too scanty to properly assess this point 
(see also section 5.3 for further discussion). 

Based on our VLT results, we find a systematic difference in the luminosity 
between RLQ and RQQ host galaxies of a factor $\sim$2 ($\sim$0.7 mag). 
Similar difference was already noted by previous studies for quasars at 
lower redshift (\citealp{veron90}; \citealp{B97}; \citealp{dunlop03}) 
and comparably high redshift \cite{Kuk01}. Whether this difference is 
intrinsic or due to some selection effect) has been long discussed 
(e.g. Hutchings, Crampton, \& Campbell 1984;  
\citealp{smith86}; \citealp{veron90}; \citealp{taylor96}; 
\citealp{hooper97}; Kirhakos et al. 1999; \citealp{dunlop03}; 
\citealp{sanchez03}). 
Main biases invoked to explain the difference are the 
non-homogeneous distribution in redshift, optical luminosity or modeling of 
the host galaxy (elliptical versus disk systems) of the compared samples. 
Our RLQ and RQQ subsamples span the same range in redshift and 
optical luminosity, therefore these effects are irrelevant. Our results, 
together with those of \citealp{B97}, \citealp{Kuk01} and \citealp{dunlop03}, 
therefore strongly indicate that the difference in host luminosity is
intrinsic and remains the same over a wide range of redshift.

For the effective radius of the host galaxies, we formally find 
$<$R$_e >$ = 10.4 $\pm$ 7.7 kpc (RLQ) and 
$<$R$_e >$ = 16.3 $\pm$ 3.4 kpc (RQQ), the reported uncertainties being the 
dispersion of the distribution, while the individual large errors have not 
been taken into account. 
The host galaxies of our high redshift quasars appear to be on average 
quite large, much larger than those found in earlier studies 
(\citealp{FKT01}; \citealp{ridgway01}) and similar to those of intermediate redshift 
RLQs (Kotilainen, Falomo \& Scarpa 1998, ; \citealp{kotifal00}).
However, given that R$_e$ is poorly constrained because of the 
degeneracy between  R$_e$  and $\mu_e$, 
we stress that the apparent difference in R$_e$ between RLQs and RQQs 
has to be considered with caution. 

\subsection{The evolution of quasar hosts}

In order to investigate the evolution of the host luminosity of RLQs and RQQs 
up to z = 2, we report in Fig. 6a the average luminosities of host galaxies 
derived from various quasar samples at z $<$ 2. We prefer to consider 
(when available) the results from HST NIR studies which are in general 
more homogeneous than those based on ground-based data.  When a sizeable 
sample is not available from HST imaging, we used NIR ground-based data. 
A further criterion in selecting data to use is that total apparent magnitudes 
of the host galaxies must be available. This is required in order to perform  
a homogeneous treatment of the data.

In the range from z = 1 to z = 2, in addition to the average values from 
this study, we use the  data from the studies by \citealp{Kuk01} and 
\citealp{ridgway01} described above. At lower redshift, we have considered 
data from the study of all RLQ hosts (34 objects) at z $<$ 0.5 imaged 
with HST \cite{pagani03}, that include previous HST studies of RLQ hosts 
by \citealp{B97}, \citealp{boyce98}, and \citealp{dunlop03}, 
and from the study of 
12 RQQs at z $\sim$0.15 by \citealp{dunlop03}. We also report the results on 
six RLQs and 10 RQQs at z $\sim$0.5 studied by \citealp{hooper97} using 
HST WFPC2 and the F675W (R) filter. Moreover, we also added the average value 
of two extensive studies of RLQ hosts in the NIR at 0.5 $<z<$ 1.0 
(\citealp{kfs98}; \citealp{kotifal00}). All these data have been made 
consistent to our system (as regards extinction, K-correction and cosmology) 
starting from the total apparent magnitudes of the host galaxies. 

We show in Figure 6 how the average host luminosities for the quasar samples 
described above evolve with redshift.  It turns out that (within the 
uncertainties of the data) the host galaxies of both types of quasars 
follow the expected passive evolution of massive ellipticals, with RLQ hosts 
being a factor of $\sim$2 more luminous than RQQ hosts. Between z = 0 and 
z = 2 there is no indication of a systematic change in the luminosity gap 
between RLQ and RQQ hosts (see also Fig. 5).  Note that both the RLQ and 
RQQ data from \citealp{hooper97} appear to lie slightly but systematically below 
(fainter hosts) the overall trend defined by the other quasar samples. 
The same trend is also apparent from comparison of their apparent magnitudes 
in the Hubble diagram (see Fig. 7).  
It is also worth to note that the increase by $\sim$0.5 mag of the 
luminosity gap between z = 1 and z = 2 proposed by \citealp{Kuk01} critically depends on one RQQ 
in their sample that is hosted by a particularly faint galaxy (see also Figure 5).  
The difference between RLQ and RQQ hosts at z$\sim$2 thus appears to be 
comparable with the gap at lower redshift within the relatively 
large uncertainty in the average values. 
In Fig. 6a, the data for the 
three RQQs studied by \citealp{ridgway01} seem to substantially deviate from 
the rest of the data. 
Apart from the possible underestimation of the host galaxy contribution 
discussed above, it is worth to note that two of the three quasars are 
significantly less luminous (by $\sim$3 mag) than the average of the other 
high redshift quasars considered here (see also next section). 
Possible effects due to differences in the nuclear luminosity cannot 
thus be excluded.

We therefore believe that the present data indicate that the difference 
between RLQ and RQQ hosts does not significantly depend on redshift, at least up to z $\sim$ 2. 
This scenario of a passive evolution of quasar hosts is consistent with 
the few available spectroscopic studies of low redshift quasar hosts and RGs 
(e.g. \citealp{cana00}; \citealp{nolan01}; \citealp{devries00}) 
indicating that their stellar content is dominated by an old well evolved 
stellar population. Finally, we note that these results do not change significantly 
if instead of the adopted cosmology we use the currently popular 
cosmology with H$_0$ = 72 km s$^{-1}$ Mpc$^{-1}$, $\Omega_m$ = 0.3 
and $\Omega_\lambda$ = 0.7 (see Figure 6b).

A  cosmic luminosity evolution, similar to that of the quasar hosts, is also displayed 
by RGs at least out to z $\sim$2.5 (Best, Longair \ \& R\"ottgering 1998;  
Lacy, Bunker, \& Ridgway 2000;  \citealp{pente01};  
\citealp{willott03}; Zirm, Dikinson, \& Dey 2003), 
whereas at even higher redshift (z $>$ 3) there is evidence that RGs have 
disturbed morphologies and a large spread in luminosity 
(Pentericci et al 1999, van Breugel et al 1999, \citealp{lacy00}). 
In Figure 7 we show the location of the RLQ hosts studied in this work, 
and the hosts of various other RLQ samples at low and high redshift, in the 
NIR apparent magnitude versus redshift diagram, 
relative to the established relation for RGs (\citealp{willott03} and 
references therein). For comparison, we also show the evolutionary model 
for elliptical galaxies derived from passive stellar evolution models 
(\citealp{bressan98}). The high-z RLQ hosts studied here fit remarkably well 
to the upper end of the RG K-z relation, better than those in \citealp{Kuk01}, 
and well within the scatter for RGs themselves. 
On the other hand, there is much larger scatter for RLQs at intermediate z, 
although the average value for RLQ hosts agrees reasonably well with 
that of the RGs. 
Both RGs and RLQs therefore appear to follow the same Hubble relation, 
indicating that both types of AGN are hosted by similar old massive 
elliptical galaxies. 

The cosmic evolution traced by quasar hosts up to z $\sim$2 disagrees with 
the expectations of semi analytic models of AGN and galaxy formation 
and evolution based on the hierarchical scenario 
(e.g. \citealp{kauffmann00}). These models  predict fainter 
(less massive) hosts at high redshift, which then merge and grow to form 
the massive spheroids observed in the present epoch. On the other hand,  
our results indicate that the luminosities of both RLQ and RQQ hosts are 
confined within a relatively small ($\sim$ 2 mag) range (see Figures 5 and 6). 
This seems in agreement with the above hierarchical model that predicts 
a small scatter in the absolute magnitude for high luminosity quasar hosts. 

Thus, if quasar hosts are luminous spheroids undergoing passive evolution, 
their mass remains essentially unchanged from z $\sim$2 up to 
the present epoch. This scenario is consistent with the results of 
recent deep surveys of distant galaxies that do not find indication of a 
drop of massive galaxies at high redshift \cite{cimatti03}. 
Alternatively, one could assume a more complex picture where 
 the mass of the hosts 
has increased from z $\sim$2 to z = 0 because of merger processes, 
as expected in the hierarchical models, and that at the same time their 
stellar content is much younger and more luminous, such as to mimic the 
passive evolution behavior. However, the observed dominance of old 
stellar population in nearby quasar host galaxies and their structural properties 
do not favor this interpretation.  Unfortunately no data are available on 
the stellar content for high redshift quasars to further asses this point.

\subsection{Nuclear versus host properties.}

Assuming that the mass of the central BH is proportional to the luminosity of 
the spheroidal component of the galaxy, as it is observed for 
nearby massive early type galaxies  
(\citealp{kor95}; \citealp{mago98}; \citealp{kor01}), and that 
the quasar is emitting at a fixed fraction of the Eddington luminosity, 
one would expect a correlation between 
the luminosity of the nucleus and that of the host galaxy. However, 
the combined effects of nuclear obscuration, possible beaming, and an 
intrinsic spread in accretion rate and 
mass-to-luminosity conversion efficiency, may destroy this correlation.

Our sample was designed to explore a broad range of nuclear luminosities 
(-25.5 $< M_B <$ -28) and can therefore be used to investigate this issue. 
In Fig. 8 we compare the $K$-band host and nuclear luminosities of 
the quasars.  While the host luminosity is distributed over a range of only 
$\sim$2 mag, the nuclear luminosities span over $\sim$4 mag. As it is apparent 
from Figure 8, there is no clear correlation between the two quantities 
in our sample. The application of the Spearman rank correlation test yields 
no correlation (R$_S$ = 0.42) with a probability of chance correlation 
p = 10\%.  
A similar negative result was also obtained from the study of lower redshift 
quasars by \citealp{dunlop03} and \citealp{pagani03}. If we supplement our 
host galaxy and nuclear dataset with those from \citealp{Kuk01} 
and \citealp{ridgway01}, we find a modest correlation 
(R$_S$ = 0.53, p $\sim$3\%). However, note that the distribution of nuclear 
and host luminosities is different for RLQs and RQQs. If the two subsamples 
are considered separately, again no significant correlation is found between 
the nuclear and host luminosities (R$_S$ $\approx$ 0.3; p $\sim$30\%). 
Some positive correlation was previously reported for lower redshift quasars 
by \citealp{hooper97}. Also in this case, however, the alleged correlation 
strongly depends on the fact that the RLQ and RQQ hosts have a significantly 
different distribution of nuclear luminosity (RQQ nuclei are $\sim$1 mag 
fainter than those of RLQs) and that the RQQ hosts are fainter than those of 
the RLQs. 

Assuming that the bolometric luminosity emitted from the 
nucleus scales as the $K$-band luminosity 
(e.g. \citealp{laor}), and that the host galaxy luminosity is proportional to 
the BH mass, it turns out that the ratio $\eta = L_{nuc}/L_{host}$ is 
proportional to the Eddington factor $\xi = L/L_E$, where 
$L_E = 1.25 \times 10^{38} \times (M_{BH}/M_{\odot}$). 
In Figure 9 we report the distribution of the $K$-band nucleus-to-host 
luminosity ratio. The average values for our full sample of 16 quasars is 
$<\log(M_{nuc}/M_{host}>_{All}$ = 1.35 $\pm$ 0.34. If RLQs and RQQs are 
considered separately an indistinguishable value is obtained: 
$<\log(M_{nuc}/M_{host}>_{RLQ}$ = 1.36 $\pm$ 0.33 and 
$<\log(M_{nuc}/M_{host}>_{RQQ}$ = 1.32 $\pm$ 0.37. 
On average our observed objects have a larger ratio nucleus/host with respect to the 
other quasars (Kukula et al.; Ridgway et al.) 
in the redshift range z=1 to 2 discussed in this work (see Figure 9).

Our results (c.f. Figures 8 and 9) therefore indicate that for 
high redshift quasars, $\xi$ is not constant, but varies in a range of 
$\Delta\xi$ $\approx$ 1.5dex. There is no significant difference in 
$\Delta\xi$ between RLQs and RQQs. A similar spread in $\xi$ was found for 
low redshift (z $<$ 0.5) RLQs (\citealp{pagani03}), suggesting that 
$\Delta\xi$ does not significantly depend on the redshift. 
This points against an interpretation of the cosmological evolution of 
quasars as being purely due to a strong luminosity evolution, 
and is more consistent with a density evolution of BH activity due to increased 
merger and fuel ling rate at high redshift. The same conclusion was reached by 
\citealp{Kuk01} on the basis of the observed modest evolution 
of quasar hosts (and their central BHs) from z = 2 to the present epoch.

\section{Summary and conclusions}

We have presented homogeneous high quality NIR images for a sample of 
17 quasars in the redshift range 1 $<$ z $<$ 2. The observations, 
obtained under excellent seeing conditions, allowed us to characterize 
the properties of the quasar host galaxies and to make a reliable comparison 
between RLQ and RQQ hosts at high redshift.

Quasar host galaxies appear to follow the same trend in luminosity of massive inactive 
ellipticals which are undergoing simple passive evolution.  There is no significant drop 
in the host mass (at least until redshift z $\sim$2) as would be (naively) 
expected in the models of joint formation and evolution of galaxies and active nuclei 
based on the hierarchical structure formation scenario. 
If this drop of mass (luminosity) occurs, it must take place at epochs earlier than z = 2. 
The same increase of host galaxy luminosity with redshift is observed both for 
RLQs and RQQs, suggesting that, in spite of their different radio properties, 
the two types of quasars are hosted by galaxies that follow the same kind 
of evolution.  However, a systematic difference in luminosity 
(and therefore likely in mass) is well apparent, indicating that RLQ hosts are 
on average a factor of $\sim$2 more luminous (massive) than RQQ hosts. 
A difference that does not appear to change significantly with the redshift.
This result appears robust since the comparison is based on samples with the same 
redshift distribution and similar nuclear luminosities. 
Nevertheless  we note that the size of the available samples of reliable 
quasar host detection at high redshift is still rather small and the objects 
do not cover the luminosity--redshift plane with adequate sampling.  
Consequently larger samples possibly covering homogeneously the full range of nuclear 
luminosity over a wide redshift interval are required to properly 
investigate if and how the nuclear luminosity contributes 
to the luminosity gap between RLQ and RQQ hosts. On this regard we also 
remark the intriguing recent result reported by Floyd et al (2003) 
who found no difference (on average) between the  host luminosity of 
RLQ and RQQ for a sample of 17 quasars at z $\sim$ 0.4 investigated 
using HST images.

The ratio between nuclear and host galaxy luminosities for the 
high redshift quasars exhibits a spread of $\sim$1.5dex. If the host galaxy 
luminosities are directly proportional to the BH mass, the observed spread 
indicates that the quasars radiate with a wide range of power with respect to 
their Eddington luminosity. The data presented here compared with the results 
of low redshift sources indicate that this spread 
does not depend on the redshift or on the radio properties of the quasars.

Since the peak epoch of quasar activity occurs at z $\sim$2.5, it will be of 
great importance to understand whether this trend exhibited by the 
quasar hosts is also followed by even higher redshift quasars. 
Exploring this issue requires the reliable characterization of the hosts of 
very distant quasars and therefore has to use facilities capable of 
high sensitivity and very narrow PSF to reduce the contribution from 
the nucleus to the extended emission. We have started a program to tackle 
this problem using VLT and NIR adaptive optics imaging 
with NACO \cite{lagrange03}.

\section*{Acknowledgments}
We thank M. Kukula and S. Ridgway for providing useful information about the 
treatment of their published data.
We also thank the anonymous referee for constructive comments and suggestions 
that improved the presentation of the results of this article.
This work was partially supported by the Italian Ministry for University 
and Research (MIUR) under COFIN 2002/27145, ASI-IR 115 and ASI-IR 35, ASI-IR 73 and by 
the Academy of Finland (project 8201017).
This publication makes use of data products from the 
Two Micron All Sky Survey, which is a joint project of the 
University of Massachusetts and the 
Infrared Processing and Analysis Center/California Institute of Technology, 
funded by the National Aeronautics and Space Administration and the 
National Science Foundation.
This research has made use of the NASA/IPAC Extragalactic Database {\em(NED)} 
which is operated by the Jet Propulsion Laboratory, 
California Institute of Technology, under contract with the 
National Aeronautics and Space Administration.

%
\begin{deluxetable}{l l l c l l l l}
\tablecolumns{7}
\tablewidth{0pc}
\tablecaption {TABLE 1}
\tablecaption  {Journal of the observations}
\tablehead{
\colhead{Quasar} & \colhead{z} & 
\colhead{V\tablenotemark{a}}& \colhead{Date} & \colhead{Filter} &
\colhead{$ T_{exp}$\tablenotemark{b}} & \colhead{Seeing\tablenotemark{c}} \\
  &  &   &   &  &  (min) & (arcsec) }
\startdata
\hline
\multicolumn{7}{c}{Radio Quiet Quasars}\\
\hline
Q0040-3731   &1.780  &17.8  &   09/June/01& K &  38 &  0.37   \\
             &       &      &   16/August/01& K &  36 &  0.56   \\
HE0935-1001  &1.574  &17.6  &   15/January/02& K &  36 &  0.46   \\
             &       &      &   15/January/02& K &  36 &  0.46   \\
             &       &      &   15/January/02& K &  36 &  0.47   \\
             &       &      &   15/January/02& K &  36 &  0.48   \\
0119-370     &1.320  &19.2  &   10/August/01& H &  30 &  0.44   \\
             &       &      &   16/August/01& H &  30 &  0.53   \\
0152-4055    &1.650  &19.3  &   09/August/01& K &  36 &  0.33   \\
             &       &      &   19/August/01& K &  36 &  0.37   \\
LBQS2135-42  &1.469  &18.35 &   30/May/01& K &  36 &  0.38  \\ 
             &       &      &   04/July/01& K &  36 &  0.33  \\ 
Q2251-2521   &1.341  &17.7  &   05/July/01& H &  36 &  0.51   \\
Q2348-4012   &1.500  &19.5  &   08/July/01& K &  36 &  0.48   \\
             &       &      &   08/July/01& K &  36 &  0.44   \\
\hline
\multicolumn{7}{c}{Radio Loud Quasars}\\
\hline
PKS0100-27 &1.597&17.8  &16/August/01& K &  36 &  0.58   \\
           &     &      &18/August/01& K &  30 &  0.39   \\
PKS0155-495 &1.298&18.4 &19/August/01& H &  36 &  0.39   \\
PKS1018-42 &1.280&18.9 &21/January/02& H &  38 &  0.33  \\ 
PKS1102-242  &1.660&19.3 & 19/January/02& K &  28 &  0.38   \\
             &     &     & 26/January/02& K &  28 &  0.38   \\
PKS1511-10  &1.513&18.5  &16/May/02& K &  36 &  0.32   \\
            &     &      &16/May/02& K &  36 &  0.36  \\ 
PKS2210-25  &1.833&19.0  &23/June/01& K &  36 &  0.37   \\
            &     &      &04/July/01& K &  36 &  0.37   \\
PKS2227-08 &1.562&17.5   &12/July/01& K &  34 &  0.34   \\
           &     &       &05/July/01& K &  30 &  0.42   \\
\enddata
\tablenotetext{a} {Quasar $V$-band magnitudes from the \citealp{veron01} catalogue;}
\tablenotetext{b} {Frame exposure time in minutes}
\tablenotetext{c} {The average FWHM in arcsec of all stars in the frame.}
\end{deluxetable}


\begin{deluxetable}{l l c l l l c }
\tablecolumns{7}
\tablewidth{0pc}
\tablecaption {TABLE 2}
\tablecaption  {Results of the radial profile modelling}
\tablehead{
\colhead{Quasar} &  \colhead{z}   &
\colhead{Filter} &  \colhead{$m_{nuc}$\tablenotemark{a}}  &
\colhead{$m_{host}\pm\Delta m_{host}$\tablenotemark{a}}
&  \colhead{r$_e\pm\Delta$r$_e$\tablenotemark{b}}    
& \colhead{$\chi^{2}_\nu(PSF)$/$\chi^{2}_\nu(Fit)$\tablenotemark{c}} \\
  &  &   &   &  & (arcsec) &   }
\startdata
\hline
\multicolumn{7}{c}{Radio Quiet Quasars}\\				 
\hline
Q0040-3731    & 1.780& K  &  15.5 &  19.4  $\pm$0.3   & (1.3)        &  6.2  \\     
HE0935-1001   & 1.574& K  &  15.1 &  $>$19.5           &              &  1.0  \\ 
0119-370      & 1.320& H  &  18.1 &  19.1$\pm$0.2     & 1.0$\pm$0.4  & 27.5  \\
0152-4055     & 1.650& K  &  17.1 &  18.6$\pm$0.3    & 1.6$\pm$1.0  & 16.0  \\
LBQS2135-42   & 1.469& K  &  15.9 &  18.7$\pm$0.4     & 1.3$\pm$1.0&  5.1  \\
Q2251-2521    & 1.341& H  &  16.0 &  18.5$\pm$0.4     & 1.6$\pm$0.7  &  9.3  \\     
Q2348-4012    & 1.500& K  &  16.9 &  19.5$\pm$0.6     & (1.1)        &  9.7  \\ 
\hline
\multicolumn{7}{c}{Radio Loud Quasars}\\	                    	 
\hline
PKS0100-27    & 1.597& K  &  15.7 &  18.5$\pm$0.5   & 0.5$\pm$0.2  &  6.5  \\     
PKS0155-495   & 1.298& H  &  17.8 &  18.4$\pm$0.2   & 0.5$\pm$0.3  & 30.7  \\     
PKS1018-42    & 1.280& H  &  15.9 &  17.6$\pm$0.3  & 1.4$\pm$0.6  & 25.2  \\ 
PKS1102-242   & 1.660& K  &  14.9 &  18.1$\pm$0.6   & (1.7)        &  5.8  \\     
PKS1511-10    & 1.513& K  &  14.8 &  18.2$\pm$0.4   & (1.8)        &  5.0  \\     
PKS2210-25    & 1.833& K  &  16.1 &  19.3$\pm$0.3   & 1.2$\pm$1.0&  2.9  \\     
PKS2227-08    & 1.562& K  &  15.8 &  18.9$\pm$0.3   & 0.3$\pm$0.2  &  3.8  \\     
\enddata
\tablenotetext{a} {Apparent magnitudes correspond to the indicated filter;}
\tablenotetext{b} {Effective radii are reported in parentheses when the 
value is uncertain due to the degeneracy of the best fit parameters.}
\tablenotetext{c} {The ratio between the reduced $\chi^2_\nu$ value of the fit with 
only the PSF model and that of the fit with PSF and host galaxy model. Only in 
the case of HE 0935-1001 the $\chi^2$ does not significantly improve 
when adding the galaxy component, therefore HE0935-1001 is indicated 
as unresolved.} 
\end{deluxetable}


\begin{deluxetable}{l l l c l l l }
\tablecolumns{7}
\tablewidth{0pc}
\tablecaption {TABLE 3}
\tablecaption  {Properties of the quasars and their host galaxies}
\tablehead{
\colhead{Quasar} &  \colhead{z}   &
\colhead{$\mu_e$\tablenotemark{a,d}} &  \colhead{K-corr\tablenotemark{b}}  &
\colhead{$M_{nucl}$\tablenotemark{c}} & \colhead{$M_{host}$\tablenotemark{c}}  & 
 \colhead{Re\tablenotemark{d}}  \\
  &  &   &   &  &   &(kpc)}
\startdata
\hline
\multicolumn{7}{c}{Radio-quiet quasars}\\
\hline
Q0040-3731      & 1.780 & (23.7)& -0.27 &  -31.5 &  -26.8  & (16)   \\     
0119-370        & 1.320 &  22.3 & 0.06  &  -28.9 &  -26.7  & 11.8   \\
0152-4055       & 1.650 &  23.4 & -0.30 &  -29.6 &  -27.3  & 20.7    \\
LBQS2135-42     & 1.469 & (22.6)& -0.32 &  -30.4 &  -26.9  & (16)   \\
Q2251-2521      & 1.341 &  22.7 & 0.07  &  -31.1 &  -27.4  & 19.6   \\     
Q2348-4012      & 1.500 & (23.7)& -0.32 &  -29.5 &  -26.1  & (13)    \\ 		
\hline
\multicolumn{7}{c}{Radio-loud quasars}\\	
\hline
PKS0000-177\tablenotemark{*} & 1.465 &  19.1 & 0.12  & -30.4  & -27.5 &  3.6   \\
PKS0100-27                   & 1.597 &  20.9 & -0.31 & -31.0  & -27.3 &  6.8   \\     
PKS0155-495                  & 1.298 &  19.9 & 0.05  & -29.1  & -27.4 &  5.7   \\     
PKS0348-120\tablenotemark{*} & 1.520 &  19.7 & 0.15  & -30.4  & -27.8 &  4.9   \\
PKS0402-362\tablenotemark{*} & 1.417 &  19.0 & 0.10  & -32.1  & -27.9 &  4.1   \\   
PKS1018-42                   & 1.280 &  21.5 & 0.04  & -31.0  & -28.1 & 16.5   \\ 
PKS1102-242                  & 1.660 & (22.9)& -0.30 & -31.9  & -27.9 & (21)   \\     
PKS1511-10                   & 1.513 & (23.3)& -0.32 & -31.7  & -27.5 & (22)   \\     
PKS2210-25                   & 1.833 & (23.3)& -0.26 & -31.0  & -27.0 & (15)   \\     
PKS2227-08                   & 1.562 &  20.0 & -0.31 & -30.7  & -26.9 &  3.7   \\   	
\enddata
\tablenotetext{*} {Results from FKT01.}
\tablenotetext{a} {Surface brightness at the effective radius (mag/arcsec$^2$) 
 derived from the best fit model.}
\tablenotetext{b} {K-correction of the host galaxy from \citealp{poggianti}.}
\tablenotetext{c} {Absolute magnitudes of the host galaxies and the nuclei 
are given in the $K$-band assuming H$_0$ = 50 km s$^{-1}$ Mpc$^{-1}$ and $\Omega$ =0.
The magnitudes are k-corrected as described in the text and no correction for galactic extinction 
is applied.}
\tablenotetext{d} {Values enclosed in parentheses are uncertain (see text).}
\end{deluxetable}


\clearpage

\begin{figure}
\plotone{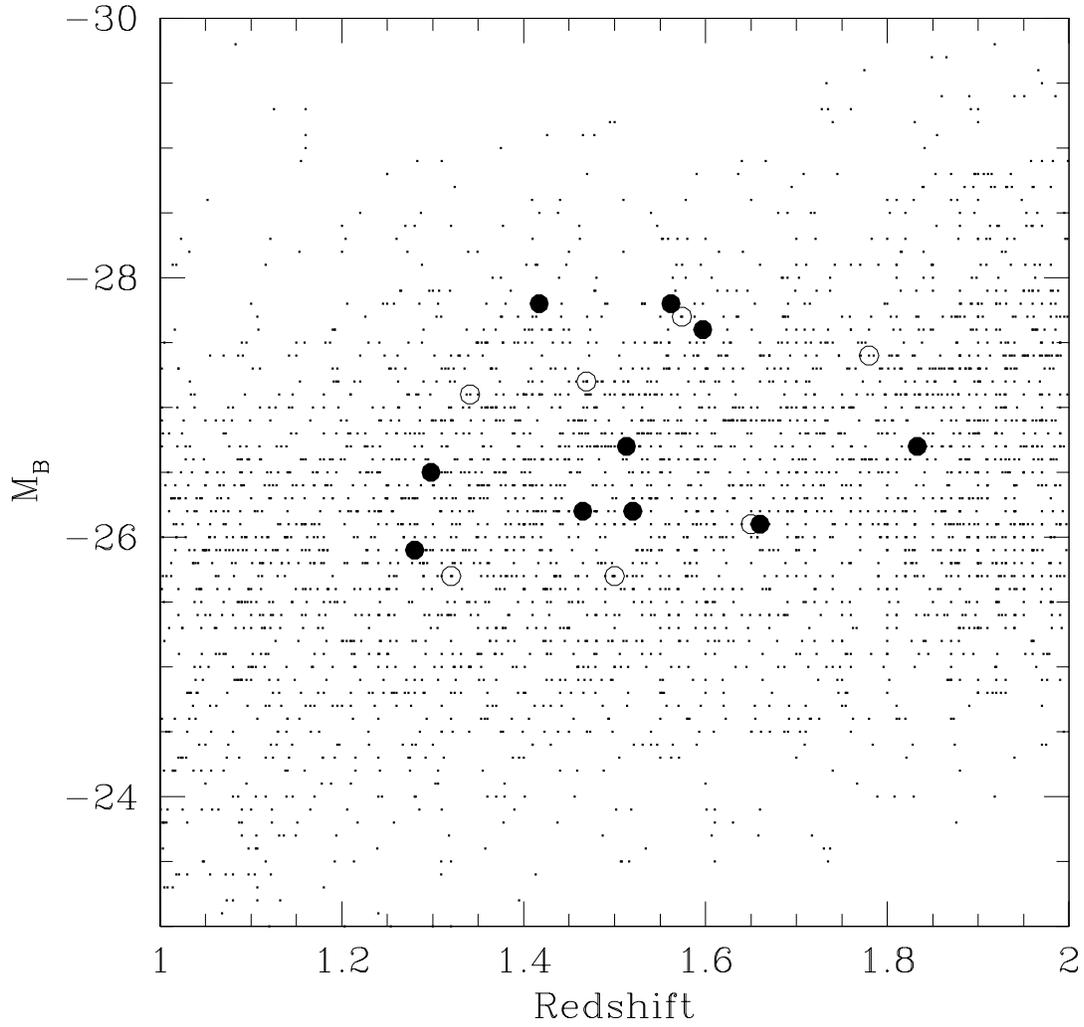}
\caption{The distribution of the observed quasars in the z - M$_B$ plane, 
compared to all quasars in the \citealp{veron01} catalogue.  The RLQs 
(filled circles) and RQQs (open circles) in our sample share an 
identical distribution in terms of redshift and optical luminosity.}
\end{figure}

\begin{figure*}
\plotone{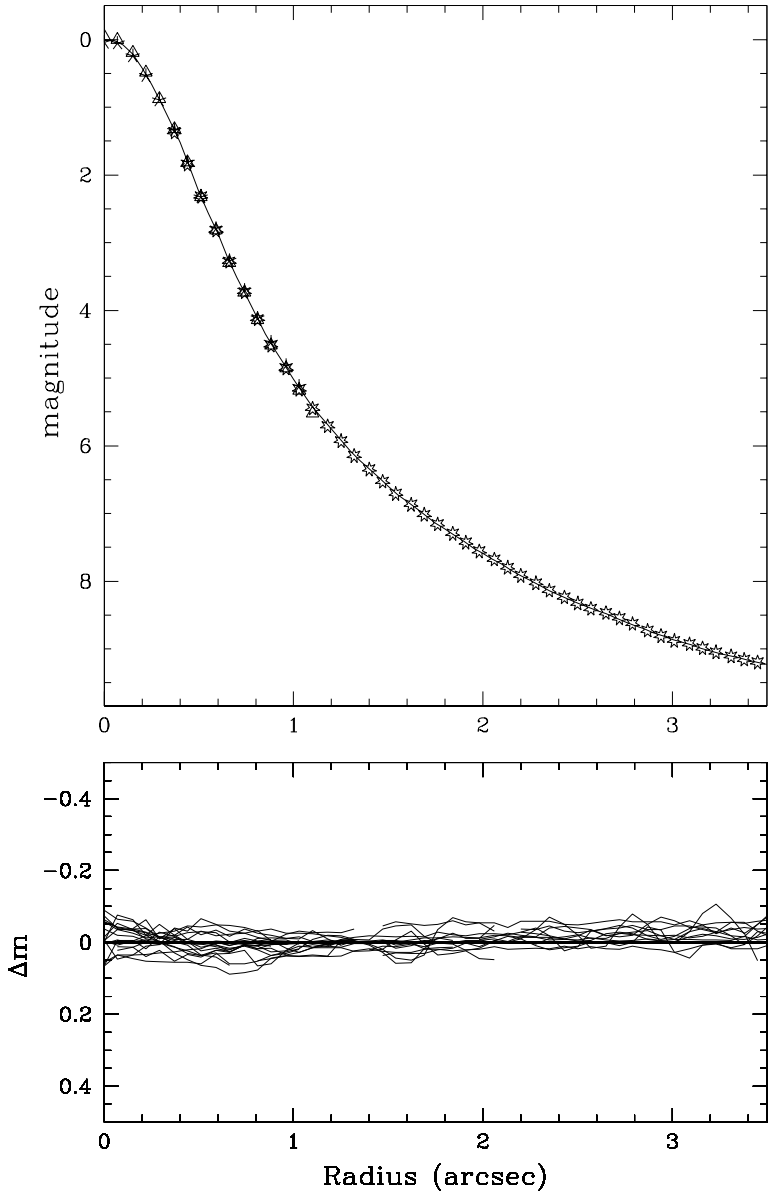}
\caption{{\it Top panel:} The PSF radial brightness profile (solid line) 
for the field of Q 2348--4012, compared with the radial profiles of 
individual stars (triangles, stars and crosses) in the frame. 
{\it Bottom panel:} Differences between the stellar radial brightness 
profiles and the adopted PSF model for all frames.}
\end{figure*}

\begin{figure*}
\plotone{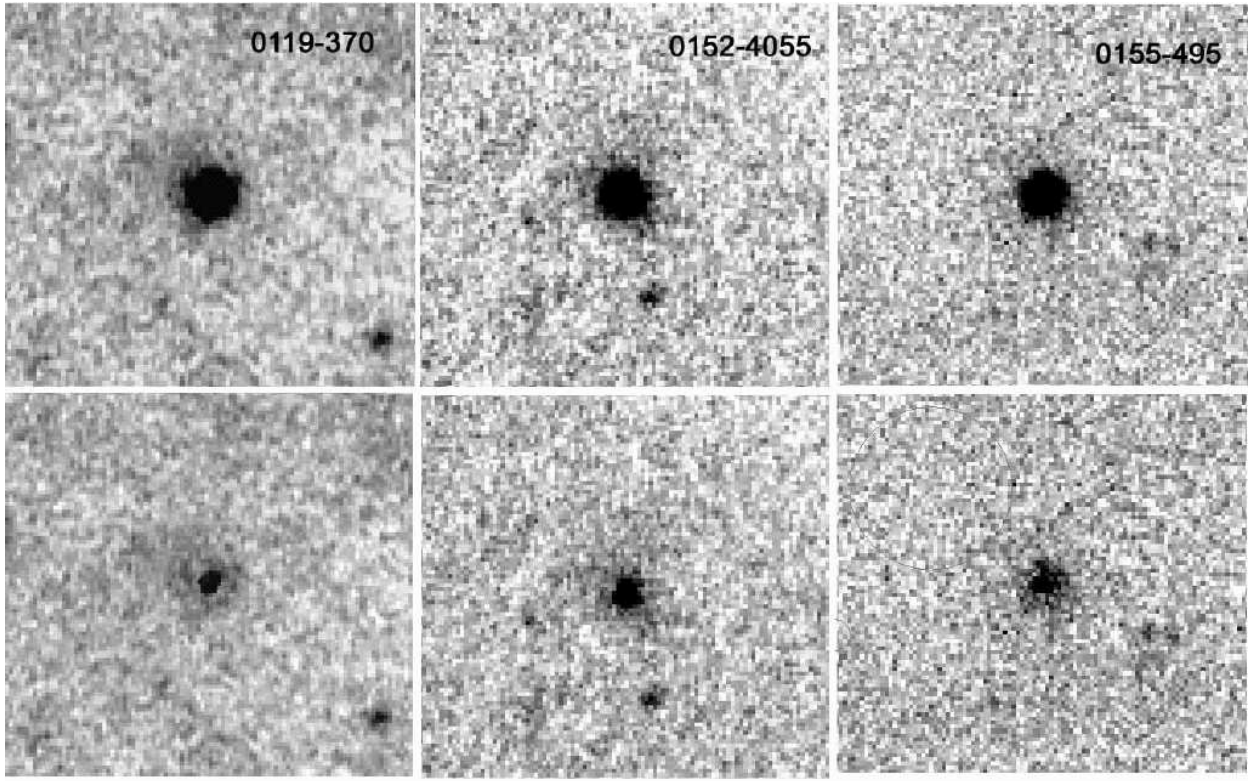}
\caption{{\it Top panel:} The near-IR images of three QSO in the sample. The 
full size of the images in each panel is of 15\asec. North is up and east to the left.
The central panel shows an example of 
 two close companion objects around a quasar that have 
 been removed in the analysis of the host galaxy (see text).  
{\it Bottom panel:} Same as above but after subtraction of a scaled PSF.}
\end{figure*}

\begin{figure*}
\plotone{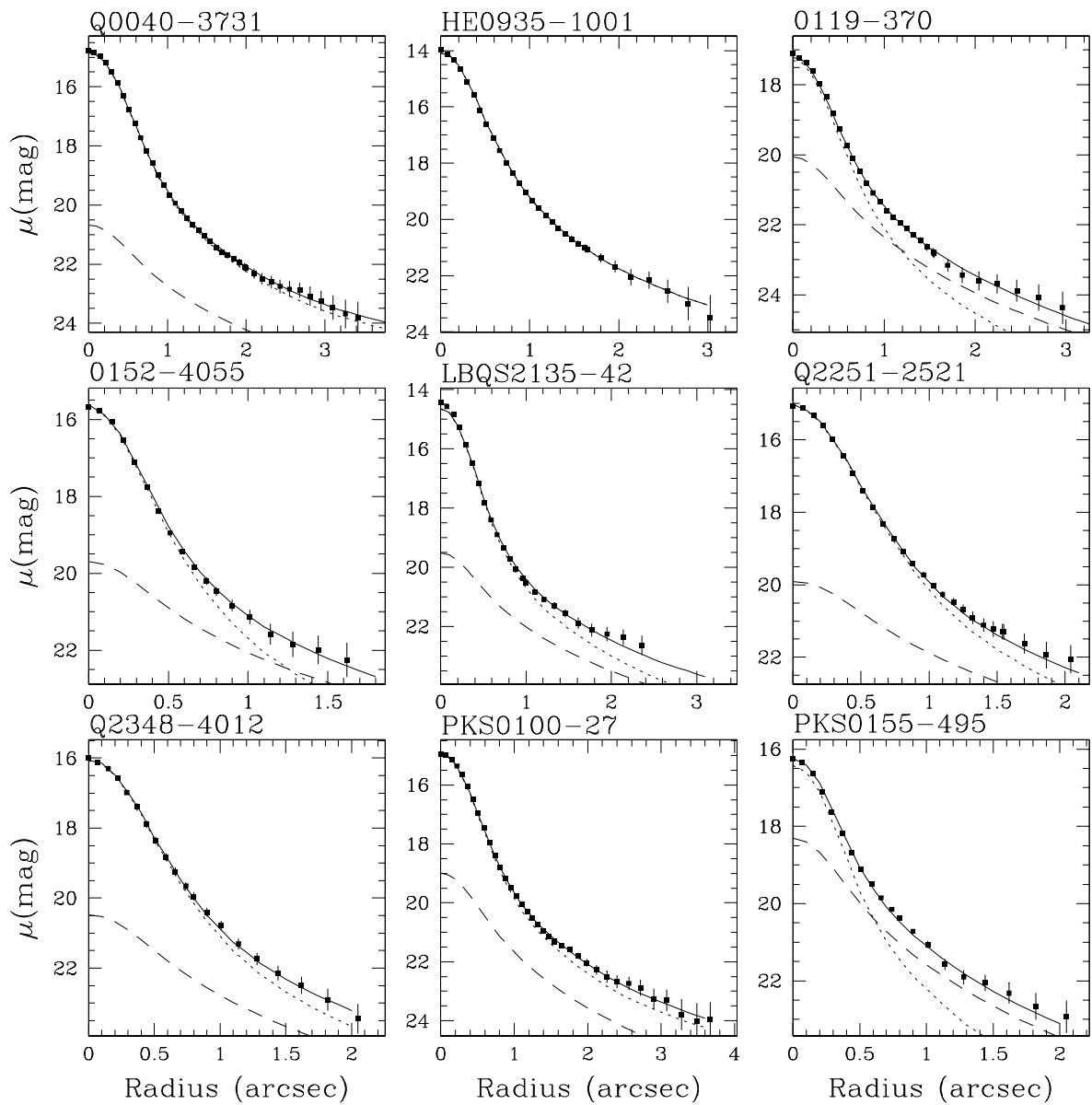}
\caption{The observed radial brightness profiles of the quasars 
(filled squares), superimposed to the fitted model consisting of the PSF 
(dotted line) and an elliptical (de Vaucouleurs law) galaxy convolved with 
its PSF (dashed line).  The solid line shows the composite model fit.}
\end{figure*}
\addtocounter{figure}{-1}%

\begin{figure*}
\plotone{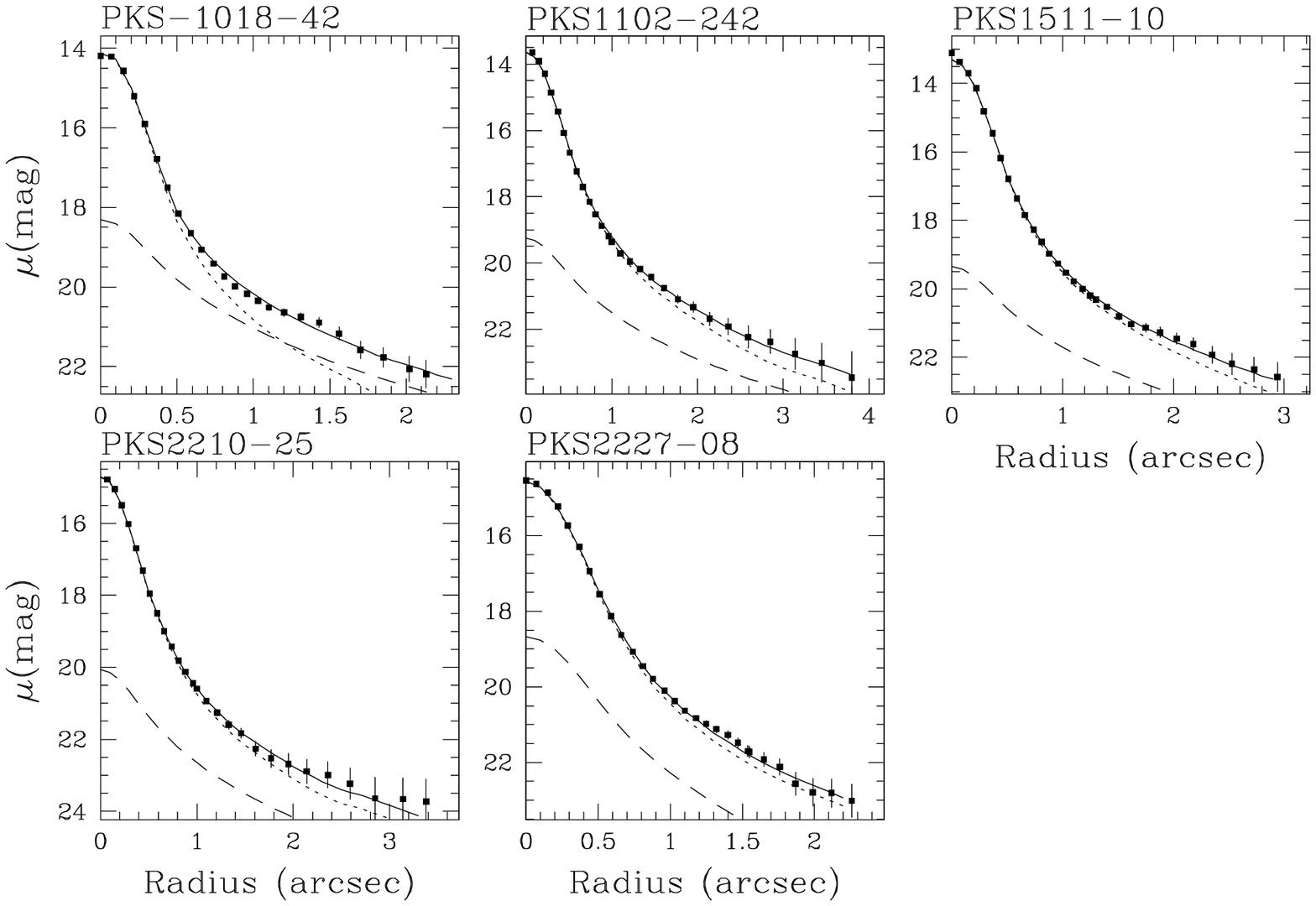}
\caption{Continued.}
\end{figure*}

\begin{figure}
\plotone{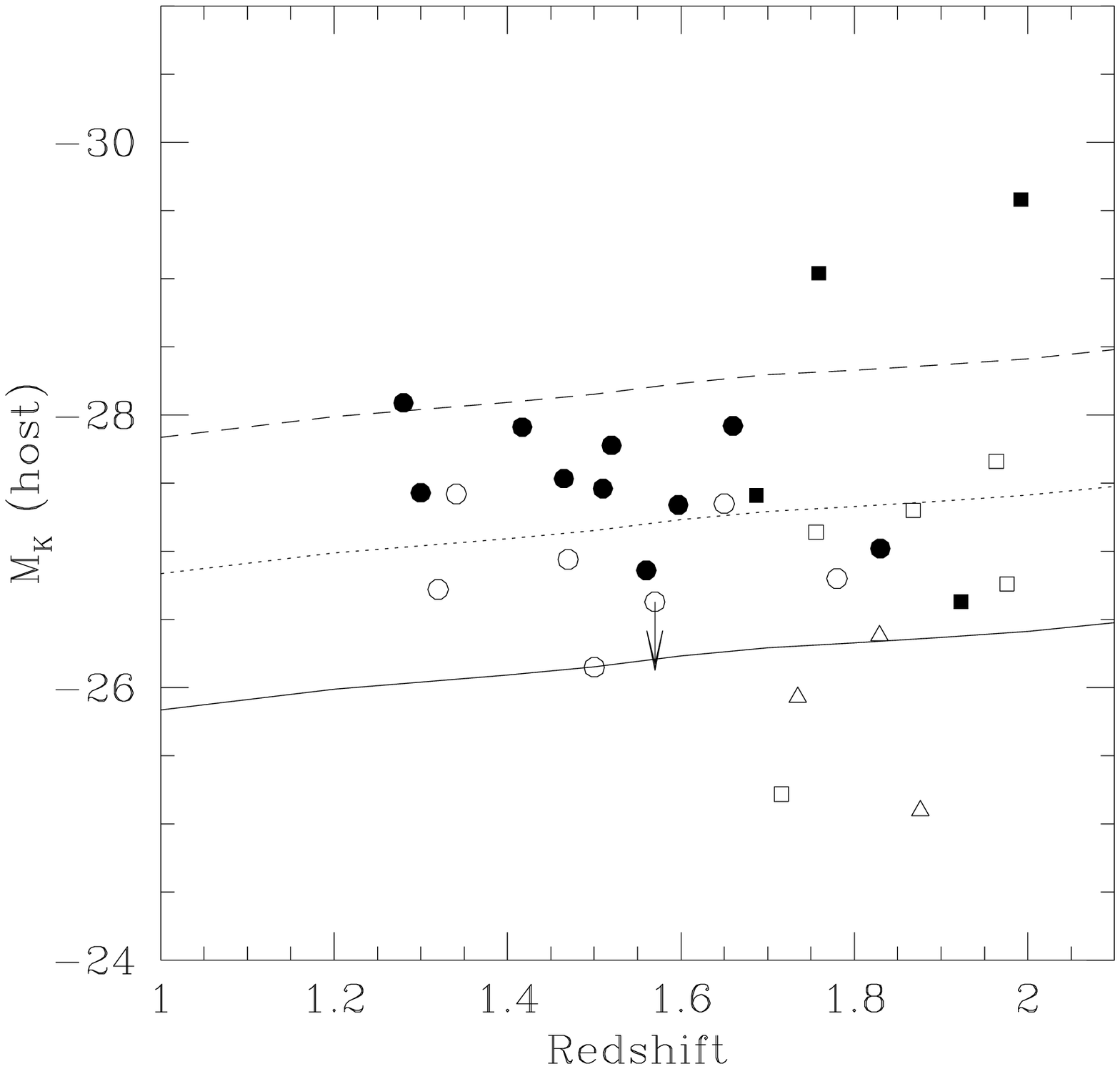}
\caption{The K band absolute magnitude of the host galaxies of
observed quasars versus redshift.  Host of RLQs (filled
circles) and RQQs (open circles) from this work are confined in the range
between M$^*$ and M$^*$ --2. 
The arrow represents the upper limit of the host luminosity 
for the unresolved object HE0935-1001.
The lines represent the expected behavior
of a massive elliptical (at M$^*$, M$^*$-1 and M$^*$-2; {\it
solid,dotted and dashed} line ) undergoing simple passive evolution \citealp{bressan98}.
Also included are the 4 RLQ (filled squares) and 5 RQQ (open squares) at
z$\sim$ 1.9 from the HST study of \citealp{Kuk01} and three RQQ (open triangles) 
from \citealp{ridgway01}.  }
\end{figure}

\begin{figure}
\plotone{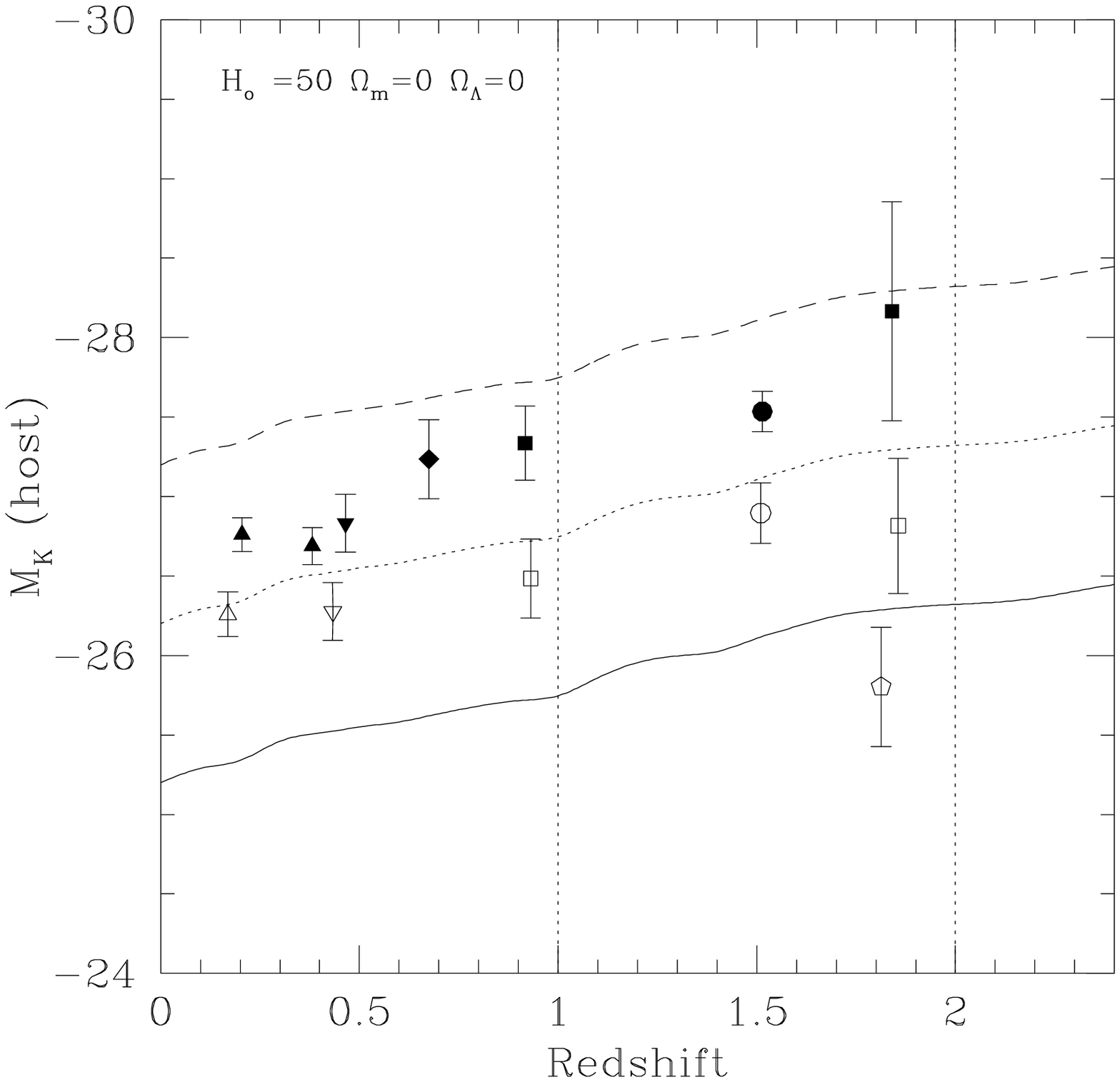}
\caption{$a)$ The evolution of quasar host luminosity compared with that expected 
for  massive ellipticals assuming H$_0$ = 50 $\Omega_m$=0 $\Omega_\Lambda$=0.
Both RLQs (filled symbols) and RQQs (open symbols) 
appear to follow the standard passive evolution for luminous 
elliptical galaxies. 
Data from this work (circles) are compared with quasar at z$\sim$0.9 and z$\sim$ 1.9 from the 
HST study of \citealp{Kuk01}  (filled squares); FSRQ and SSRQ study at z$\sim$0.8 from 
\citealp{kfs98} and \citealp{FKT01} 
(filled diamonds); 
low redshift RLQs compiled from HST observations \cite{pagani03} (filled triangles) 
divided in two bins (objects at redshift smaller and larger than z=0.25); 
low redshift RQQ from 
\citealp{dunlop03} (open triangles) and 3 RQQ at z $\sim$ 1.8 (open pentagons) by 
\citealp{ridgway01}; data for z $\sim$ 0.5 are taken from \citealp{hooper97} 
(inverted filled triangle for RLQ and inverted open triangles for RQQ).
Each point is plotted at  the mean redshift of the considered 
sample while the error bar represents the 
dispersion of the mean value of the host luminosity.
The lines represent the expected behavior
of a massive elliptical (at M$^*$, M$^*$-1 and M$^*$-2; {\it
solid,dotted and dashed} line ) undergoing passive stellar 
evolution \citealp{bressan98}. 
$b)$ Same as $a)$ but using the cosmology H=50 $\Omega_m$ = 0.3 $\Omega_\Lambda$ = 0.7.
}
\end{figure}
\addtocounter{figure}{-1}%

\begin{figure}
\plotone{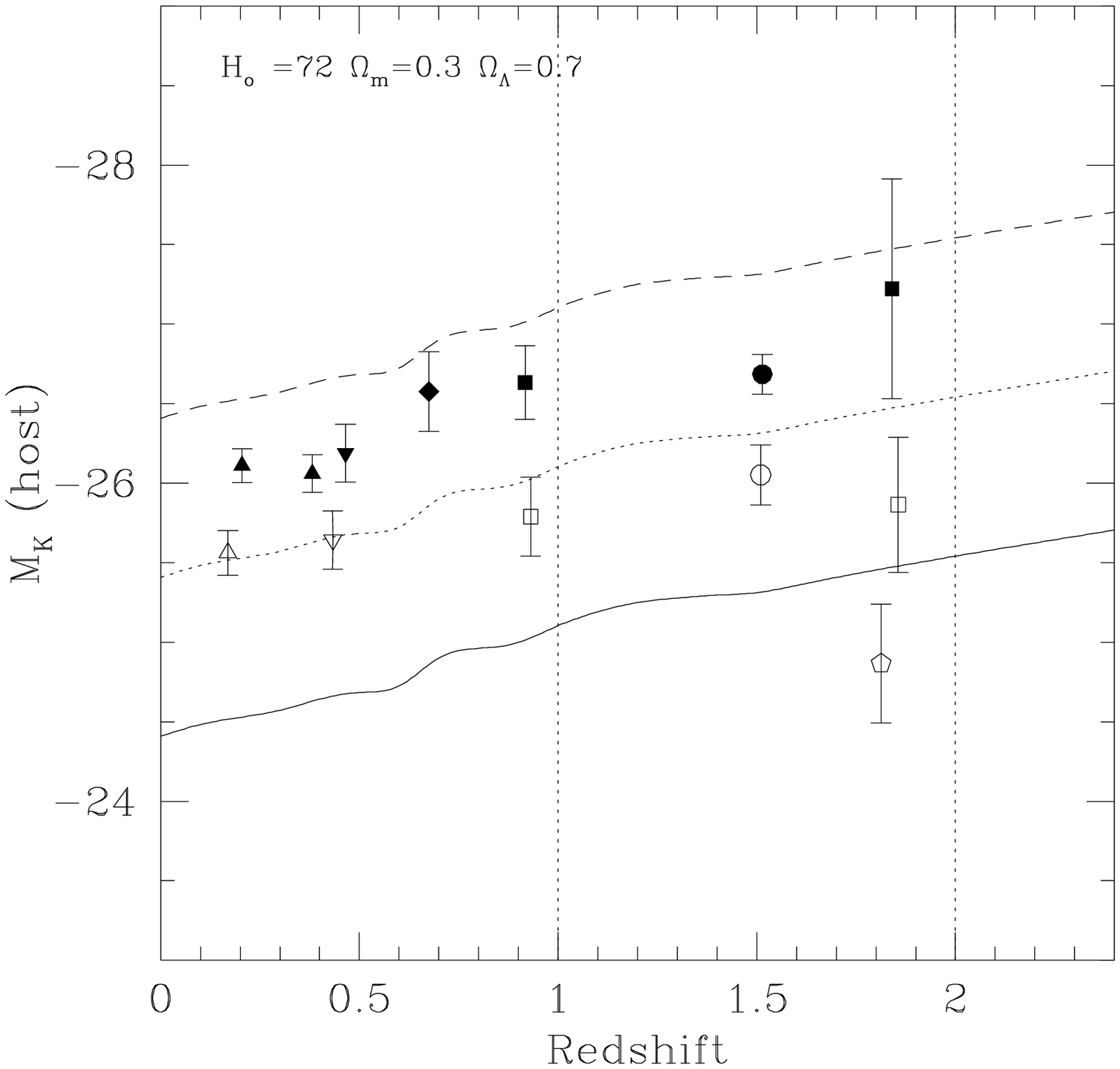}
\end{figure}

\begin{figure}
\plotone{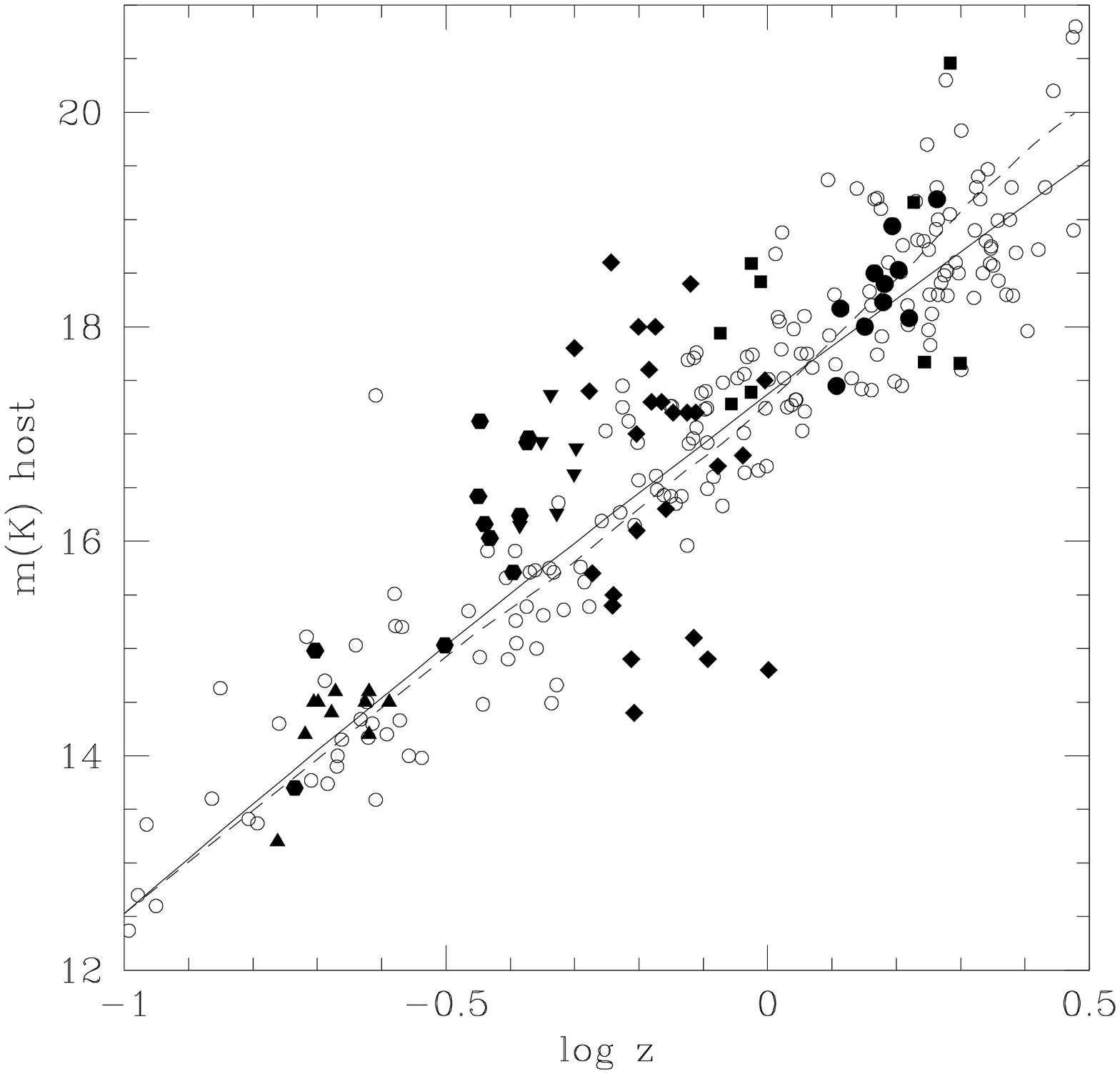}
\caption{
The K-band apparent magnitude of the host galaxies versus the redshift for 
RLQ samples at high redshift (this work, filled circles; 
\citealp{Kuk01}, filled squares), intermediate redshift objects 
(Kotilainen et al. 1998, \citealp{kotifal00}, filled diamonds; 
Hooper et al. 1997; filled inverted triangles) and low redshift 
(\citealp{pagani03}, filled hexagons; \citealp{dunlop03}, 
filled triangles). Also shown are data for RGs (\citealp{willott03}, 
open circles), the best-fit  relationship for 
RGs (\citealp{willott03}, solid line) and the model of passive evolution 
(\citealp{bressan98}, dashed line).
}
\end{figure}

\begin{figure}
\plotone{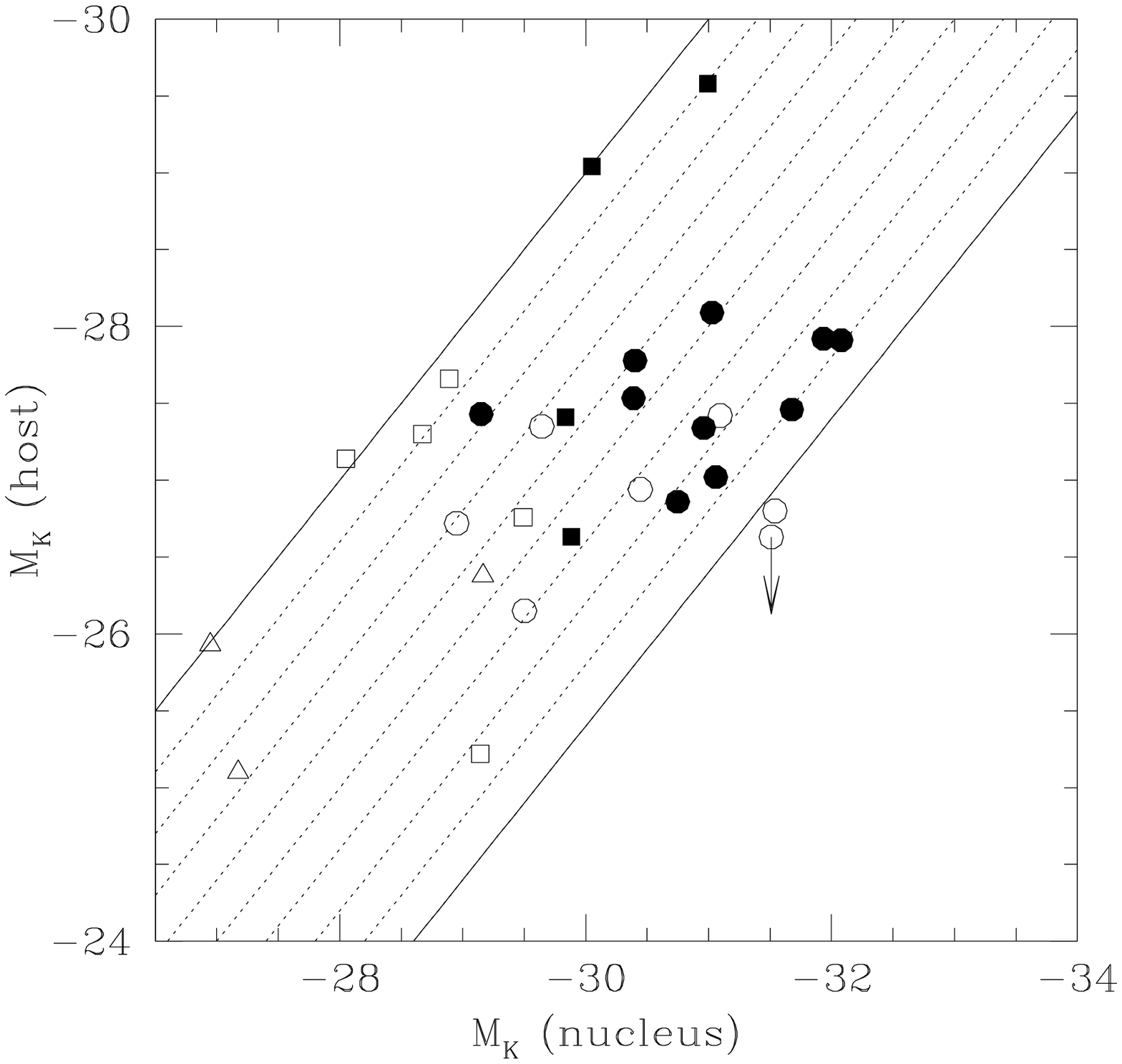}
\caption{The absolute magnitude of the nucleus compared with that of the host galaxy.
RLQ (filled circles) and RQQ (open circles) spanning a  similar range 
of nuclear luminosity are hosted in galaxy of $\sim$ 1 magnitude different while 
nuclear luminosity ranges about 3 magnitudes. 
The arrow represents the upper limit of the host luminosity derived for the unresolved object HE0935-1001.
No significant correlation is found.
For comparison also data from  \citealp{Kuk01} (squares) 
and \citealp{ridgway01} (triangles) are reported. 
Diagonal lines represent the loci of constant ratio between host and nuclear 
emission. These can be translated into Eddington 
ratios assuming that the central BH mass - galaxy luminosity correlation 
is hold up to z $\sim$ 2 and that the observed nuclear 
power is proportional to the bolometric emission. 
Separations between dotted lines correspond to a 
difference by a factor of 2 in the nucleus-to-host luminosity ratio.
The two solid lines encompass a spread of 1.5dex in this ratio.
}
\end{figure}

\begin{figure}
\plotone{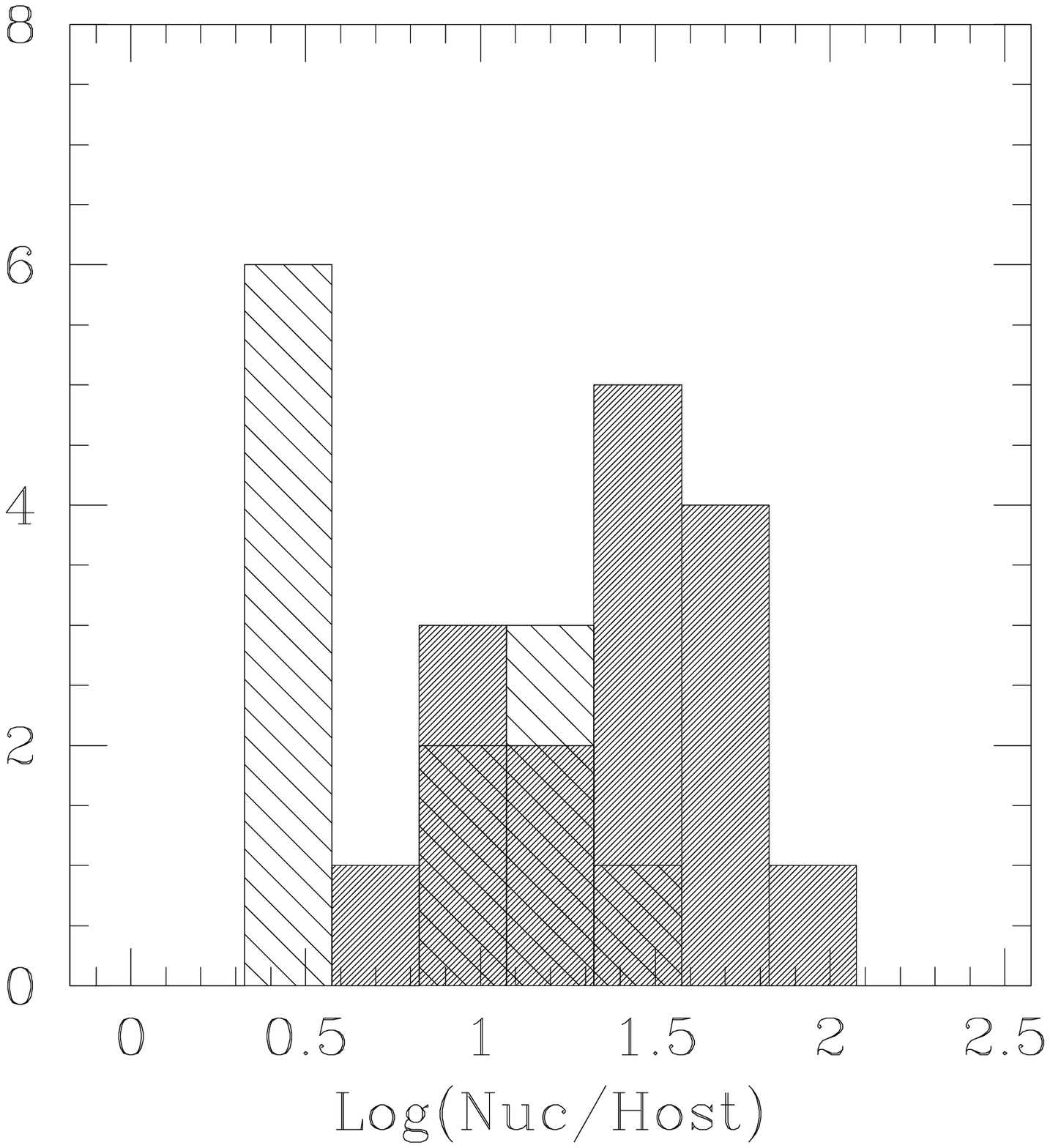}
\caption{The distribution of the nuclear to host luminosity ratio $\eta$ 
for the whole sample of high redshift quasars observed at VLT (shaded area) compared with that 
of the 12 objects studied by Kukula et al and Ridgway et al.
Our sample show on average an higher nuclear-to-host luminosity ratio ($<$Log(nuc/host)$>$ = 1.3) 
with respect to the the other quasars considered in this study ($<$Log(nuc/host)$>$ = 0.8) .
For the VLT sample no significant difference of nucleus/host ratio is found between 
RLQ and RQQ which also exhibit a similar spread ( $\sim$ 1.5dex ) of the nucleus/host ratio. 
}
\end{figure}


\begin{thebibliography}{}

\bibitem [Abraham Crawford \& McHardy 1992]{abraham92} Abraham R.G., Crawford C.S., McHardy I.M., 1992, ApJ, 401, 474
\bibitem [Abraham et al. 1996]{abraham96} Abraham, R.G., van den Bergh, S., Glazebrook, K., Ellis, R.S., Santiago, B.X., Surma, P., Griffiths, R.E. 1996, ApJS, 107, 1
\bibitem [Aretxaga et al. 1998]{arex98} Aretxaga, I., Terlevich, R.J., Boyle, B.J. 1998, MNRAS, 296, 643 
\bibitem [Bahcall et al. 1997]{B97} Bahcall, J.N., Kirhakos S., Saxe D.H., Schneider D.P. 1997, ApJ 479, 642
\bibitem [Best et al. 1998] {best98} Best, P.N., Longair, M.S., R\"ottgering, H.J.A., 1998, MNRAS 295, 549
\bibitem [Boyce et al. 1998]{boyce98} Boyce, P.J., Disney, M.J., Blades, J.C., Boksenberg, A., Crane, P., Deharveng, J.M., Macchetto, F.D., Mackay, C.D., Sparks, W.B. 1998, MNRAS, 298, 121
\bibitem [Boyle \& Terlevich 1998]{boyle98}  Boyle, B. J., Terlevich, R.J. 1998 MNRAS, 293, L49
\bibitem [Boyle et al. 2000]{boyle00} Boyle, B.J., Shanks, T., Croom, S.M., Smith, R.J., Miller, L., Loaring, N., Heymans, C. 2000,
MNRAS, 317, 1014
\bibitem [Boyle 2001]{boyle01} Boyle, B.J. 2001, Advanced Lectures on the Starburst-AGN Connection, 
(ed. I.Aretxaga, D.Kunth, R.Mujica), Singapore: World Scientific, p.325
\bibitem  [Bressan Granato \& Silva 1998]{bressan98} Bressan, A., Granato G.L. \& Silva L. 1998, A\&A 332, 135.
\bibitem [Canalizo \& Stockton 2000]{cana00} Canalizo, G., Stockton, A., 2000, ApJ 528, 201 
\bibitem [Cimatti 2003]{cimatti03} Cimatti A. 2003 astro-ph/0303023.
\bibitem [Cuby et al. 2000]{cuby} Cuby, J.G., Lidman, C., Moutou, C., Petr, M. 2000, Proc. SPIE, 4008, 1036
\bibitem [Devillard 1999]{devillard99} Devillard, N., 1999, Astronomical Data Analysis Software and Systems VIII, ASP Conference Series, Vol. 172 (ed. D.M. Mehringer, R.L. Plante, D.A. Roberts), p. 333
\bibitem [de Vries et al. 2000]{devries00} de Vries, W.H, O'Dea, C.P, Barthel, P.D., Fanti, C, Fanti, R, Lehnert, M.D., 2000, AJ 120, 2300
\bibitem [Disney et al. 1995]{disney95} Disney, M.J., Boyce, P.J., Blades, J.C., Boksenberg, A., Cane, P., Deharveng, J.M., Macchetto, F., Mackay, C. D., Sparks, W.B., Phillipps, S. 1995, Nature, 376, 150
\bibitem [Dunlop \&  Peacock 1990]{dunlop90}  Dunlop, J. S., Peacock, J. A. 1990 MNRAS, 247,19
\bibitem [Dunlop et al. 2003]{dunlop03} Dunlop, J.S., McLure, R.J., Kukula, M.J., Baum, S.A., O'Dea C.P., Hughes, D.H. 2003, MNRAS, 340, 1095
\bibitem [FKT01]{FKT01} Falomo, R. Kotilainen, J.K., Treves, A. 2001, ApJ 547 124 (FKT01)
\bibitem [Falomo Carangelo \& Treves 2003]{falomo03} Falomo, R. Carangelo, N. \& Treves, A. 2003, MNRAS 343, 505 
\bibitem [Ferrarese 2002]{ferrarese02} Ferrarese, L., 2002, Proceedings of the 2nd KIAS Astrophysics Workshop, 
(ed. C.-H.Lee, H.-Y.Chang) Singapore: World Scientific Publishing, p.3
\bibitem[Floyd et al 2003] {floyd03} Floyd D.J.E., Kukula, M.J., Dunlop, J.S., McLure, R.J., Miller, L., Percival, W.J., Baum, S.A., O'Dea, C.P.
\bibitem [Franceschini et al. 1999] {franc99} Franceschini, A., Hasinger G., Miyaji T., Malquori D. 1999, MNRAS, 310, L5 
\bibitem [Francis et al. 2000 ] {francis00} Francis, P.J. Whiting, M.T. \& Webster, R.L. 2000 PASA 17, 56
\bibitem[Govoni et al. 2000]{govoni00} Govoni, F., Falomo, R., Fasano, G., \& 
Scarpa, R. 2000a, A\&A, 353, 507
\bibitem[Jorgensen et al. 1996]{jorgensen96} J\o rgensen, I., Franx, M., \& Kj\ae rgaard, P. 1996, MNRAS, 280, 167
\bibitem [Hamilton Casertano \& Turnshek 2002] {hamilton02} Hamilton T.S., Casertano S. Turnshek D.A. 2002 ApJ, 576, 61
\bibitem [Heckman et al. 1991]{heckman} Heckman, T.M., Miley, G.K., Lehnert, M.D., van Breugel, W. 1991 ApJ, 381, 373
\bibitem [Hooper et al. 1997]{hooper97} Hooper, E.J., Impey C.D. \& Foltz C.B., 1997, ApJ, 480, L95
\bibitem [Hutchings et al. 1984]{hutchings84} Hutchings, J.B., Crampton, D. \& Campbell, B. 1984 ApJ, 280, 41
\bibitem [Hutchings \& Morris 1995]{hutch95}  Hutchings, J. B., Morris, S. C. AJ 109, 1541
\bibitem [Hutchings 1998]{hutch98} Hutchings, J. B. 1998 AJ, 116, 20
\bibitem [Hutchings et al. 1999] {hutch99} Hutchings, J. B., Crampton, D., Morris, S.L., Durand, D., Steinbring, E. 1999, AJ, 117, 1109 
\bibitem [Kauffmann \& Haehnelt 2000]{kauffmann00} Kauffmann, G., Haehnelt, M., 2000, MNRAS, 311, 576
\bibitem [Kauffmann et al. 2003]{kauffmann03} Kauffmann, G., et al. 2003, MNRAS 341, 54
\bibitem [Kinney et al. 1996] {kinney96} Kinney, A.L., Calzetti, D., Bohlin, R.C., McQuade, K., 
Storchi-Bergmann, T. \& Schmitt, H.R. 1996, ApJ 467 38
\bibitem [Koo et al. 1996]{koo} Koo, D.C., Vogt, N.P., Phillips, A.C., Guzman, R., Wu, K.L., Faber, S.M., Gronwall, C., Forbes, D.A.,
Illingworth, G.D., Groth, E.J., Davis, M., Kron, R.G., Szalay, A.S. 1996, ApJ, 469, 535
\bibitem [Kormendy \& Richstone 1995]{kor95} Kormendy, J., Richstone, D. ARA\&A, 33, 581
\bibitem [Kormendy \& Gebhardt 2001]{kor01} Kormendy, J. \& Gebhardt, K. 2001, AIP conference proceedings, 586, 363
\bibitem [Kotilainen Falomo \& Scarpa 1998]{kfs98}Kotilainen, J.K., Falomo, R., Scarpa, R. 1998,  A\&A, 332, 503 
\bibitem [Kotilainen \& Falomo 2000]{kotifal00} Kotilainen, J.K., Falomo, R. 2000,  A\&A, 364, 70
\bibitem [Kukula et al. 2001]{Kuk01} Kukula, M.J., Dunlop, J.S., McLure, R.J., Miller, L., Percival, W.J., Baum, S.A., O'Dea, C.P. 2001, MNRAS, 326, 1533
\bibitem [Lacy et al. 2000]{lacy00} Lacy, M., Bunker, A.J., Ridgway, S.E., 2000, AJ, 120, 68
\bibitem [Lacy et al. 2001]{lacy01} Lacy, M., Laurent-Muehleisen, S.A., Ridgway, S.E., Becker, R.H., White, R.L. 2001, ApJ, 551, L17
\bibitem [Lagrange et al. 2003]{lagrange03} Lagrange, A.-M., Chauvin, G., Fusco, T., et al., 2003, SPIE 4841, 860
\bibitem [Laor \& Draine 1993]{laor} Laor, A., Draine, B.T. 1993,  ApJ, 402, 441
\bibitem [Le Fevre et al. 2000]{lefevre} Le Fevre, O., Abraham, R., Lilly, S.J., Ellis, R.S., Brinchmann, J., Schade, D., Tresse, L.,
Colless, M., Crampton, D., Glazebrook, K., Hammer, F., Broadhurst, T. 2000, MNRAS, 311, 565
\bibitem [Lehnert et al. 1992]{lehnert92} Lehnert, M.D., Heckman, T.M., Chambers, K.C., Miley, G.K. 1992 ApJ, 393, 68
\bibitem [Lehnert et al. 1999]{lehnert99} Lehnert, M.D., van Breugel, W.J.M., Heckman, T.M., Miley, G.K. 1999 ApJS, 124, 11
\bibitem [Lowenthal et al. 1995]{lowenthal} Lowenthal, J.D., Heckman, T.M., Lehnert, M.D., Elias, J.H. 1995 ApJ, 439, 588
\bibitem [Madau et al. 1998]{madau98} Madau, P., Pozzetti, L., Dickinson, M. 1998 ApJ, 498, 106 
\bibitem [Magorrian et al. 1998]{mago98} Magorrian, J.,Tremaine, S., Richstone, D., Bender, R., Bower, G., Dressler, A., Faber, S.M.,
Gebhardt, K., Green, R., Grillmair, C., Kormendy, J., Lauer, T. 1998, AJ, 115, 2285
\bibitem [McLeod \& Rieke 1994]{mcleod94} McLeod, K.K., Rieke, G.H., 1994, ApJ, 431, 137
\bibitem [Mobasher et al. 1993]{mob93} Mobasher, B.; Sharples, R. M.; Ellis, R. S. 1993, MNRAS 263, 560
\bibitem [Nolan et al. 2001]{nolan01} Nolan, L.A, Dunlop, J.S, Kukula, M.J, Hughes, D.H, Boroson, T, Jimenez, R., 2001, MNRAS 323, 308
\bibitem [Pagani et al. 2003]{pagani03} Pagani, C., Falomo, R., Treves, A., 2003, 
ApJ 596, 830
\bibitem [Pentericci et al. 2001]{pente01} Pentericci ,L., McCarthy, P.J., R\"ottgering, H.J.A., Miley, G.K., van  Breugel, W.J.M., Fosbury, R., 2001, ApJS 135, 63
\bibitem [Percival et al. 2000]{perci00} Percival, W.J., Miller, L., McLure, R.J., Dunlop, J.S. 2000, MNRAS 322, 843
\bibitem [Poggianti 1997]{poggianti}Poggianti, B.M., 1997, A\&AS, 122, 399
\bibitem [Ridgway et al. 2001]{ridgway01} Ridgway, S., Heckman, T., Calzetti, D., Lehnert, M. 2001, ApJ, 550, 122
\bibitem[Sanchez \&  Gonzalez-Serrano] {sanchez03} 2003 A\&A 406, 435
\bibitem [Smith et al. 1996] {smith86} Smith, E.P. Heckman, T.M. Bothun, G.D. 
Romanishin, \& W. Balick, B. 1986 ApJ, 306, 64
\bibitem [Steidel et al. 1999]{steidel99} Steidel, C.C., Adelberger, K.L., Giavalisco, M., Dickinson, M., Pettini, M., 1999, ApJ 519, 1
\bibitem [Taylor et al. 1996]{taylor96} Taylor, G.L., Dunlop, J.S., Hughes, D.H., Robson, E.I. 1996, MNRAS, 283, 930
\bibitem [Veron-Cetty \& Woltjer 1990]{veron90} Veron-Cetty, M.P., Woltjer, L. 1990, A\&A, 236, 69
\bibitem [Veron-Cetty \& Veron 2001] {veron01} Veron-Cetty, M.P., Veron, P. 2001, A\&A, 374, 92
\bibitem [Warren et al. 1994]{warren94} Warren, S.J., Hewett, P.C., Osmer, P.S. 1994, ApJ, 421, 412W 
\bibitem [Willott et al. 2003]{willott03} Willott, C.J., Rawlings, S., Jarvis, M.J, Blundell K.M. 2003, MNRAS 339, 173
\bibitem [Zirm et al. 2003] {zirm03} Zirm, A.W., Dikinson, M. \& Dey, A. 2003, ApJ, 595, 90 
\end{thebibliography}
\end{document}